\documentclass[twocolumn]{aastex701}

\usepackage{amsmath}
\usepackage{comment}
\usepackage{multirow}

\newcommand{\Mdot}{{\rm M}_{\odot}}
\newcommand{\logMdot}{\log~({\rm M}_{\ast}/{\rm M}_{\odot})}
\newcommand{\logcharMass}{\log~({\rm M}^{*}/{\rm M}_{\odot})}
\newcommand{\kms}{{\rm km}~{\rm s}^{-1}}
\newcommand{\dn}{D_{n}4000}

\newcommand{\angstrom}{\text{\AA}}

\begin{document}

\title{The Stellar Mass Function for Nine Massive Galaxy Clusters in the Local Universe} 

\correspondingauthor{Jubee Sohn} 
\email{jubee.sohn@snu.ac.kr} 

\author[0009-0009-4334-5598]{Jong-In Park}
\affiliation{Astronomy Program, Department of Physics and Astronomy, Seoul National University, 1 Gwanak-ro, Gwanak-gu, Seoul 08826, Republic of Korea}
\email{jongin.park@snu.ac.kr}

\author[0000-0002-9254-144X]{Jubee Sohn}
\affiliation{Astronomy Program, Department of Physics and Astronomy, Seoul National University, 1 Gwanak-ro, Gwanak-gu, Seoul 08826, Republic of Korea}
\affiliation{SNU Astronomy Research Center, Seoul National University, Seoul 08826, Republic of Korea}
\email{jubee.sohn@snu.ac.kr}

\author[0000-0002-9146-4876]{Margaret J. Geller}
\affiliation{Smithsonian Astrophysical Observatory, 60 Garden Street, Cambridge, MA 02138, USA}
\email{mgeller@cfa.harvard.edu}

\author[0000-0001-5014-8108]{Ken J. Rines}
\affiliation{Department of Physics and Astronomy, Western Washington University, Bellingham, WA 98225, USA}
\email{rinesk@wwu.edu}

\author{Antonaldo Diaferio}
\affiliation{Università di Torino, Dipartimento di Fisica, Torino, Italy}
\affiliation{Istituto Nazionale di Fisica Nucleare (INFN), Sezione di Torino, Torino, Italy}
\affiliation{Accademia delle Scienze di Torino, via Maria Vittoria 3, 10123 Torino, Italy}
\email{diaferio@ph.unito.it}

\begin{abstract}
We measure galaxy stellar mass functions (SMFs) for nine of the most massive galaxy clusters in the local universe ($0.07 < z < 0.11$) using deep and complete spectroscopy from the MAssive Cluster Survey with Hectospec (MACH). We construct the cluster SMFs down to $\logMdot \gtrsim 8.5$. For comparison, we measure the SMF for field galaxies, complete to $\logMdot \gtrsim 10.5$, based on Sloan Digital Sky Survey (SDSS) spectroscopy over the same redshift range. The mean MACH SMF shows a shape similar to that of the field SMF but with a significantly higher amplitude at $\logMdot < 11.4$. At $\logMdot > 11.4$, the MACH SMF shows a clear excess, indicating the contribution of massive galaxies, including Brightest Cluster Galaxies (BCGs). Based on homogeneous MACH spectroscopy, we compare SMF shapes for quiescent and star-forming members as a function of cluster-centric distance. The quiescent SMFs display a curved shape with a peak at $\logMdot \approx 10.5$; the star-forming SMFs decline monotonically with increasing stellar mass. We further compare the mean MACH SMF with SMFs derived from similarly massive clusters in the IllustrisTNG-300 simulations. The shape of the observed and simulated SMFs agree well overall. However, the MACH clusters contain roughly a factor of two more galaxies at $9.0 < \logMdot < 10.5$. These results demonstrate that constructing cluster SMFs from complete spectroscopic samples can test simulations and provide powerful constraints on galaxy formation and evolution in dense environments.
\end{abstract}

\keywords{Galaxy clusters (584) --- Stellar mass function (1616) --- Spectroscopic surveys (1558)}

\section{Introduction}

The dense environments of galaxy clusters play a crucial role in the formation and evolution of massive galaxies. Cluster galaxies provide a well-defined sample for statistical studies of galaxy evolution at high density. Comparing statistical distributions of various galaxy properties for cluster and field galaxies offers valuable constraints on the environmental processes that shape galaxy evolution.

Stellar mass is one of the primary physical parameters governing galaxy evolution (e.g., \citealp{Peng2010}). Stellar mass correlates strongly with a wide range of other galaxy properties, including stellar velocity dispersion, size, star formation rate, and metallicity (e.g., \citealp{Kauffmann2003b, Tremonti2004, Noeske2007, Damjanov2009, Williams2010, Zahid2016, Sohn2017}); it also exhibits some dependence on environment and local density (e.g., \citealp{Kauffmann2004, Bolzonella2010, Vulcani2012}). The stellar mass function (SMF, hereafter), the number density of galaxies as a function of stellar mass, provides a fundamental statistical framework for studying galaxy formation and its dependence on environment.

In the standard structure formation paradigm, galaxies form within dark matter (DM) (sub)halos; DM halos first accrete baryons, and subsequent baryonic processes shape galaxy properties. Comparisons between the theoretical DM halo mass function and the observed SMF have long been a critical tools for constraining the role of baryonic physics and the efficiency of galaxy formation (e.g., \citealp{Kauffmann1993, Benson2002, Bower2006, Croton2006, Somerville2008, Behroozi2010, Behroozi2018}). Discrepancies between the two have motivated the introduction of various feedback mechanisms (e.g., \citealp{Ciotti1997, Silk1998, Menci2002, Shankar2006, Hopkins2014}). 

Large spectroscopic surveys have enabled systematic derivations of the SMF in different redshift ranges (e.g., \citealp{Peng2010, Mortlock2011, Baldry2012, Vulcani2013, Weigel2016, Driver2022, Xu2025, ArizoBorillo2025}). For example, \citet{Weigel2016} measured the field SMF using SDSS Data Release (DR) 7 for $\logMdot \gtrsim 9$ at $0.02 < z < 0.06$, and \citet{Driver2022} derived the field SMF for $\logMdot \gtrsim 9$ at $z < 0.1$ using the Galaxy And Mass Assembly (GAMA) DR4 survey. More recently, \citet{Xu2025} used the Dark Energy Spectroscopic Instrument (DESI) Bright Galaxy Survey (BGS) to derive the SMF for $z < 0.2$. They extend their determination to very low stellar masses ($\logMdot \sim 6$) by combining spectroscopy with photometric redshifts for fainter galaxies.

Robust SMF measurements in massive clusters based on deep and complete spectroscopy remain relatively rare (e.g., \citealp{Balogh1999, Vulcani2011, Vulcani2013, Sohn2017, vanderBurg2018, vanderBurg2020}). For example, \citet{Vulcani2011} derived cluster SMFs using the WIde-field Nearby Galaxy cluster Survey (WINGS; $0.04 < z < 0.07$) and the ESO Distant Cluster Survey (EDisCS; $0.4 < z < 0.8$), covering stellar masses $\logMdot \gtrsim 10$. \citet{Vulcani2013} extended this work to $0.3 < z < 0.8$ with the IMACS Cluster Building Survey (ICBS; $0.3 < z < 0.45$) and EDisCS. They focused on the high-mass end ($\logMdot \gtrsim 10.5$) based on nearly 4000 member galaxies within 18 clusters. At higher redshift, \citet{vanderBurg2018, vanderBurg2020} measured SMFs for clusters at $0.5 < z < 0.8$ and $1.2 < z < 1.5$ using the GOGREEN survey \citep{Balogh2017}.

Deep and complete spectroscopy of nearby massive clusters enables investigation of SMFs at lower stellar masses. For example, \citet{Ferrarese2016} measured the SMF for the Virgo cluster even including globular clusters. They utilize the Next Generation Virgo Survey (NGVS) covering the central region of the Virgo cluster. Later, \citet{Morgan2025} derived the Virgo SMF down to ${\logMdot} \sim 7.8$ within a wider spatial area (${\sim 104~\mathrm{deg}^2}$) based on the NGVS and the Virgo Environmental Survey Tracing Ionised Gas Emission (VESTIGE). \citet{Sohn2017} derived SMFs for Coma and Abell 2029 for $\logMdot > 8$. These measurements, comparable in mass range to field SMF determinations, provide a critical baseline for understanding galaxy formation as a function of local density.

Here, we derive the SMFs of nine local massive galaxy clusters at a redshift range of $0.07 < z < 0.11$ and a virial mass range of $5.5 \times 10^{14} \, \mathrm{M}_{\odot} \lesssim M_{200} \lesssim 10^{15} \, \mathrm{M}_{\odot}$, based on deep and highly complete spectroscopy from the MACH survey (MAssive Cluster surveys with Hectospec) with MMT/Hectospec. A purely magnitude-limited survey without color-based target selection, MACH enables unbiased SMF measurements for both quiescent and star-forming populations the same limiting stellar mass. Based on this dataset, we carefully determine cluster membership and stellar masses. We then compare the resulting cluster SMFs with field SMFs at matched redshift ranges. We further compare the results with predictions from cosmological simulations, providing a test of galaxy formation models in dense environments.

We describe the spectroscopic survey dataset in Section \ref{sec:data}. Section \ref{sec:cluster_smf} explains the construction of the cluster SMFs. In Section \ref{sec:comparison_cluster_field}, we derive the field SMF in the redshift range of the MACH clusters based on SDSS spectroscopy. We investigate the SMFs of quiescent and star-forming galaxies and compare the MACH SMFs with those derived from the IllustrisTNG-300 cluster simulations in Section \ref{sec:discussion}. We conclude in Section \ref{sec:conclusion}. We adopt a standard $\Lambda$CDM cosmology with $\Omega_m = 0.3$, $\Omega_{\Lambda} = 0.7$, and $H_{0} = 70~\text{km s}^{-1} \text{Mpc}^{-1}$ throughout the paper.

\section{DATA} \label{sec:data}

\subsection{The MACH Cluster Sample} \label{sec:mach_data}

The Massive Cluster Survey with Hectospec (MACH, \citealp{Sohn2020}) is a spectroscopic survey of the nine most massive clusters in the local universe ($0.07 < z < 0.11$) using MMT/Hectospec. MACH is a dense and complete spectroscopic survey, including more than 4500 spectra for each target cluster. Table \ref{tab:tab1_MACHinfo} summarizes the physical properties of the clusters and the survey characteristics for the 9 MACH clusters. 

\begin{figure}[t]
\centering
\includegraphics[width=0.5\textwidth]{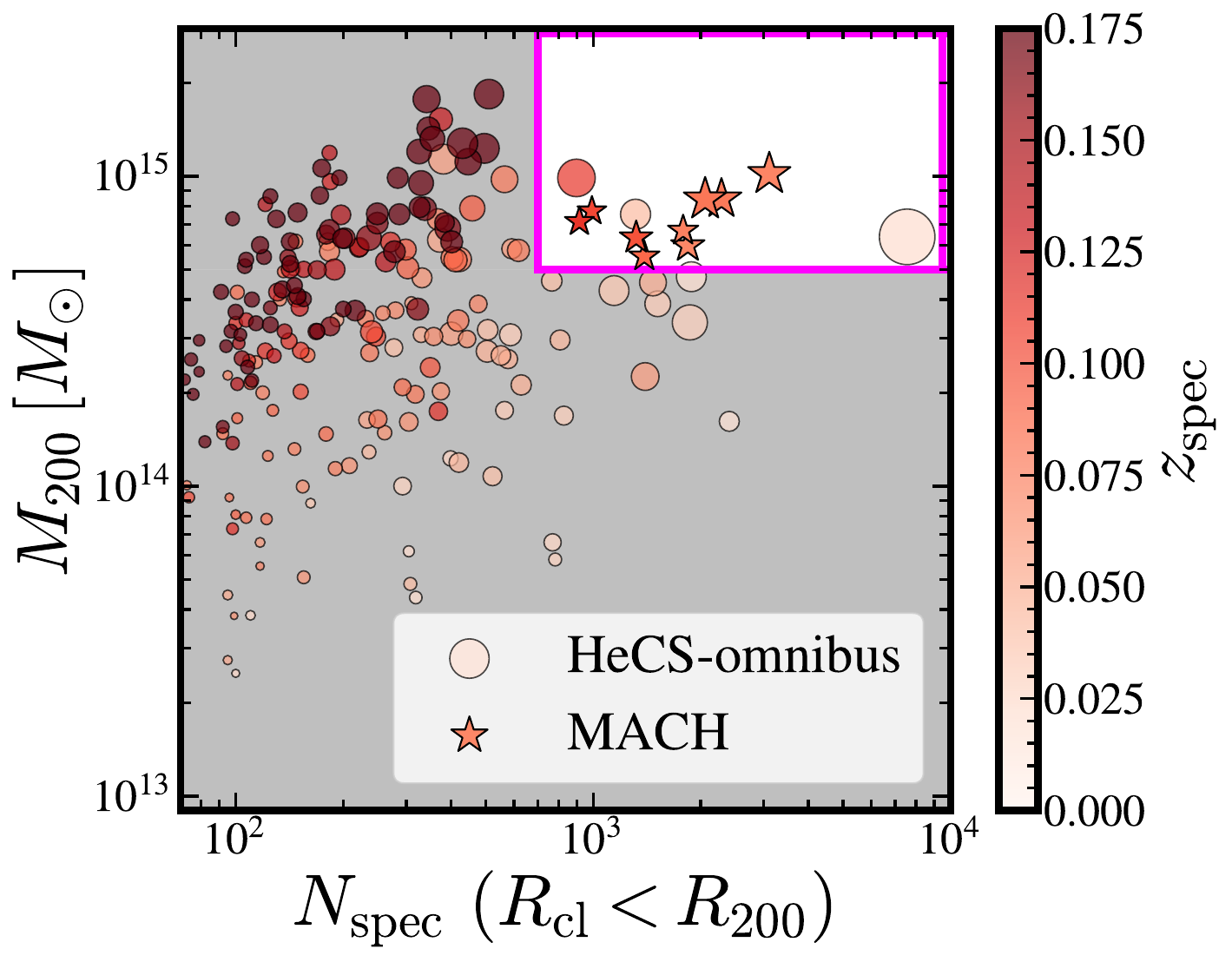}
\caption{$M_{200}$ of galaxy clusters in the HeCS-omnibus sample \citep{Sohn2020} as a function of the number of galaxies with spectroscopy within each cluster field. Stars and circles indicate the MACH clusters (this study) and other clusters in the HeCS-omnibus sample \citep{Sohn2020}, respectively. Redder symbols indicate clusters at higher redshifts. The size of the symbol is proportional to the number of spectroscopically identified cluster members within $R_{200}$.}
\label{fig:mach_info}
\end{figure}

Figure \ref{fig:mach_info} shows the cluster mass (i.e., $M_{200}$) and the number of galaxies with spectra ($N_{\rm spec}$) in each cluster field. Here, $M_{200}$ is the cluster mass enclosed within $R_{200}$ where the total mass density corresponds to 200 times the critical density at the cluster redshift. We derive $R_{200}$ and $M_{200}$ based on the caustic technique (\citealp{Diaferio1997, Diaferio1999, Serra2013}, see Section \ref{sec:cluster_membership}). The $N_{\rm spec}$ is the count of galaxies with spectra and with $R_{\rm cl} < R_{200}$, where $R_{\rm cl}$ is the projected cluster-centric distance. 

In Figure \ref{fig:mach_info}, circles show the HeCS-omnibus clusters \citep{Sohn2020} with $0.01 < z < 0.30$ and stars show the nine MACH clusters with $0.07 < z < 0.11$. Symbols are color-coded based on the cluster redshift; the size of the symbol indicates the number of spectroscopically identified members within $R_{\rm cl} < R_{200}$. The magenta square marks the region where MACH clusters are located. MACH clusters are the most massive and well-sampled clusters among the HeCS-omnibus clusters. The mass range of the MACH clusters range is $5.52 \times 10^{14} \Mdot < M_{200} < 1.02 \times 10^{15} \Mdot $. The typical mass is $7.4 \times 10^{14} \Mdot$.  

Three clusters in the magenta box are not included in the MACH sample: Coma ($z = 0.023$), A119 ($z = 0.044$) and A2069 ($z = 0.114$). These clusters are not included in the MACH sample because they are outside the sample redshift range. The main purpose of the MACH survey is to provide the dense spectroscopy for the most massive clusters within the redshift range $0.07 < z < 0.11$. We select the redshift range $0.07 < z < 0.11$ to match the Hectospec aperture. The alignment with the Hectospec aperture allows us to construct deep spectroscopy efficiently.

\subsubsection{Photometric Data}

\begin{deluxetable*}{lccccccc}
\tablewidth{\textwidth}
\label{tab:tab1_MACHinfo}
\setlength{\tabcolsep}{10pt}
\caption{MACH Clusters}
\tablehead{
\colhead{ID\tablenotemark{a}} & \colhead{R.A. (deg)} & \colhead{Decl. (deg)} & \colhead{Redshift} & \colhead{$N_{\text{spec}}$} & \colhead{$N_{\text{spec}}(<3R_{200})$} & \colhead{$N_{\text{spec}}(<R_{200})$} & \colhead{$r_{\text{comp, limit}}$\tablenotemark{b}}}
\startdata
A2245 & 255.676 & 33.525 & 0.088 & 39270 & 7543 & 1390 & 20.875 \\
A1767 & 204.038 & 59.188 & 0.071 & 13394 & 4733 & 1840 & 20.875 \\
A2244 & 255.677 & 34.060 & 0.099 & 39129 & 6577 & 1319 & 20.875 \\
A1831 & 209.810 & 27.989 & 0.075 & 18466 & 5265 & 1785 & 20.125 \\
A2034 & 227.551 & 33.520 & 0.113 & 37944 & 4273 & 915  & 20.625 \\
A7    & 2.937   & 32.417 & 0.103 & 13463 & 4144 & 992  & 20.875 \\
A2255 & 258.101 & 64.017 & 0.080 & 11944 & 5087 & 2057 & 20.625 \\
A2029 & 227.729 & 5.767  & 0.079 & 34829 & 8506 & 2235 & 20.625 \\
A2065 & 230.601 & 27.697 & 0.073 & 25159 & 7954 & 3109 & 21.125 \\ \hline \hline
\enddata
\tablenotetext{a}{We order the clusters based on their characteristic mass, $M_{200}$ (see Table \ref{tab:tab2_MACHcausticinfo}).}
\tablenotetext{b}{The completeness limit ($r_{\text{comp, limit}}$) is defined as the magnitude where the cumulative spectroscopic completeness drops below 90\%, evaluated within a bin width of 0.125.}
\end{deluxetable*}

\setcounter{footnote}{0}

We obtain photometric catalogs for the MACH clusters from Sloan Digital Sky Survey (SDSS) Data Release 18 (DR18, \citealp{Almeida2023}). We select galaxies with \texttt{probPSF} = 0, where \texttt{probPSF} indicates the probability of being a point source. We also select galaxies brighter than an extinction-corrected Petrosian magnitude of 22 in $r-$band. For the initial construction of the photometric catalog, we include sources within a wide field with a radius of $180^\prime$. For MACH clusters, this field typically covers an area with a radius larger than $7R_{200}$. We also investigated SDSS photometric objects with unreliable photometry based on the photometric flags, including objects with saturated pixels, fatal photometric errors, failed profile fitting, deblended objects without distinct peaks, and contamination by cosmic rays. For these objects, we visually inspected them based on SDSS color images to exclude artifacts, including stellar bleeding, fragmentation of bright galaxies. We excluded 930 artifacts from the visual inspection.

We use the SDSS $ugriz$ photometry for analyses including stellar mass estimation. Among various photometric magnitudes, we use the composite model (cModel) magnitudes\footnote{The composite model magnitude is a measurement of galaxy brightness that combines the best aspects of exponential and de Vaucouleurs profile fits. An optimal linear combination of the two models allows accurate representation of galaxies that are not well described by a single profile type and provides a smooth transition between disk-dominated and bulge-dominated systems.}. We apply the foreground extinction correction to all magnitudes. Hereafter, all photometric magnitudes indicate foreground extinction corrected cModel magnitude (e.g., $r_{\rm cModel, 0}$) unless noted otherwise. 

We note that the SDSS DR18 photometric catalog is incomplete over a small patch in the A2029 field \citep{Sohn2017}; for this area, we compile SDSS DR7 photometry \citep{Abazajian2009} to recover missing objects. Additionally, we replace SDSS DR18 photometry for galaxies with poor measurements. For example, the BCG of A2029, i.e., IC 1101, one of the brightest galaxies in the local universe, has poor photometry in DR18; the cModel magnitude of IC 1101 is only $r = 17.41$. For this object, we use the SDSS DR7 photometry for IC 1101, which is more reliable ($r = 13.46$). There are also 14 objects in all MACH cluster fields that we identify based on visual inspection of the SDSS images with reasonable DR7 photometry and improper DR18 photometry. We use the DR7 photometry for these objects. 

In A2245 and A7, there are some areas where the survey seems incomplete (see Figure \ref{fig:2Dspec_completeness} for details). In these areas, bright saturated stars prevent construction of complete photometric catalogs. We note, however, that the impact of saturated stars is negligible because the areas affected by those stars are tiny (i.e., less than 0.5\% of the area within $R_{200}$).

\subsubsection{Spectroscopy}

We use new spectroscopic redshifts for MACH clusters to determine cluster membership and to estimate the stellar masses of galaxies. We conduct a deep and dense spectroscopic survey with MMT/Hectospec. We also compile redshifts from previous surveys and from the  literature, including SDSS DR18  \citep{Almeida2023} and DESI DR1 \citep{AbdulKarim2025}.  

MACH is a deep and complete spectroscopic survey with Hectospec \citep{Fabricant2005}, a 300-fiber-fed spectrograph mounted on the 6.5m MMT. The one-degree diameter field of view enables efficient surveys of rich clusters. The Hectospec spectra cover the wavelength range $3750 - 9100~\angstrom$ with a resolution of $R \sim 1500$. 

We conducted the Hectospec survey for MACH clusters from 2017 to 2023. We selected galaxies brighter than $r_\mathrm{cModel, 0} = 21.3$ as spectroscopic targets. We excluded low surface brightness objects with $r$-band fiber magnitudes fainter than 22 mag. We did not impose any color selection. The exposure time of the Hectospec observations is an hour ($3 \times 1200$ second exposures). We obtained a total of 41897 unique spectra for objects in nine MACH cluster fields ($\sim 4500$ objects per cluster). 

We use HSRED v2.0 \footnote{http://oirsa.cfa.harvard.edu/archive/search/} to reduce the Hectospec spectra. We derive spectroscopic redshifts based on RVSNUpy \citep{Kim2025}. We select redshifts with a cross correlation score $R_{XC} > 3$ following \citet{Rines2016}. The typical redshift uncertainty is $c\Delta z \sim 19.2~\kms$. 

The Sloan Digital Sky Survey (SDSS) provides spectroscopy for bright galaxies in the MACH cluster fields. The SDSS spectra cover the wavelength range $3800 - 9200~{\rm \angstrom}$ with a spectral resolution of $R \sim 2000$. The SDSS spectroscopy is nearly complete for galaxies with $r \leq 17.77$ mag. There are 59700 unique SDSS redshifts in the nine MACH cluster fields.

We also collect spectroscopic redshifts from the Dark Energy Spectroscopic Instrument (DESI) survey \citep{AbdulKarim2025}. The DESI spectroscopy covers a wavelength range $3600 - 9800~\angstrom$ with a spectral resolution of $R \sim 2000 - 5000$. We identify 175760 objects with DESI redshifts within $180\arcmin$ of the centers of the nine MACH clusters; 16859 of these are located within $3R_{200}$. Finally, we compile 372 spectroscopic redshifts for the nine MACH clusters from the NASA/IPAC Extragalactic Database (NED, see Table \ref{tab:NED_redshift} in Appendix \ref{sec:NED_database}).

\begin{figure*}[htbp]
\centering
\includegraphics[width=\textwidth]{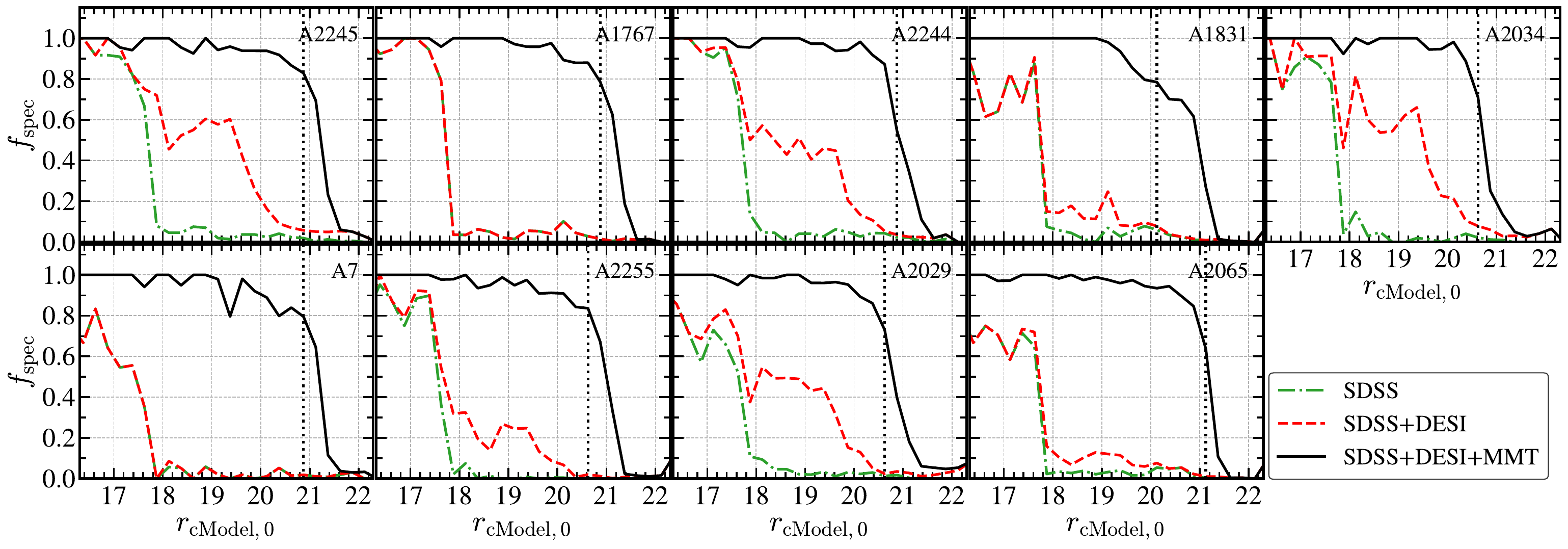}
\caption{Spectroscopic survey completeness for galaxies within each cluster field (i.e., $R_\mathrm{cl} < R_{200}$), as a function of the foreground extinction-corrected $r$-band composite model magnitude. The black vertical dotted lines indicate a cumulative completeness of 90\% for each cluster. The $M_{200}$ of the clusters increases from left to right and from top to bottom.}
\label{fig:1Dspec_completeness} 
\end{figure*}

Figure \ref{fig:1Dspec_completeness} illustrates the differential spectroscopic survey completeness ($f_\mathrm{spec}$) as a function of $r$-band magnitude for galaxies within $R_{200}$ in each cluster field. We define $f_\mathrm{spec}$ as the ratio between the number of spectroscopic objects and the number of photometric objects in each $r$-band magnitude bin. In Figure \ref{fig:1Dspec_completeness}, we plot the $f_\mathrm{spec}$ computed from samples based only on SDSS (green dot-dashed lines), SDSS and DESI (red dashed lines), and the entire spectroscopic dataset including the Hectospec surveys (black solid lines). Figure \ref{fig:1Dspec_completeness} highlights the impressive impact of the dense Hectospec survey on the the survey completeness. The MACH spectroscopic sample is more than 90\% complete for $r < 20.5$; for A1831 the sample is 90\% complete for $r < 20.2$. 

\begin{figure*}[htbp]
\centering
\includegraphics[width=\textwidth]{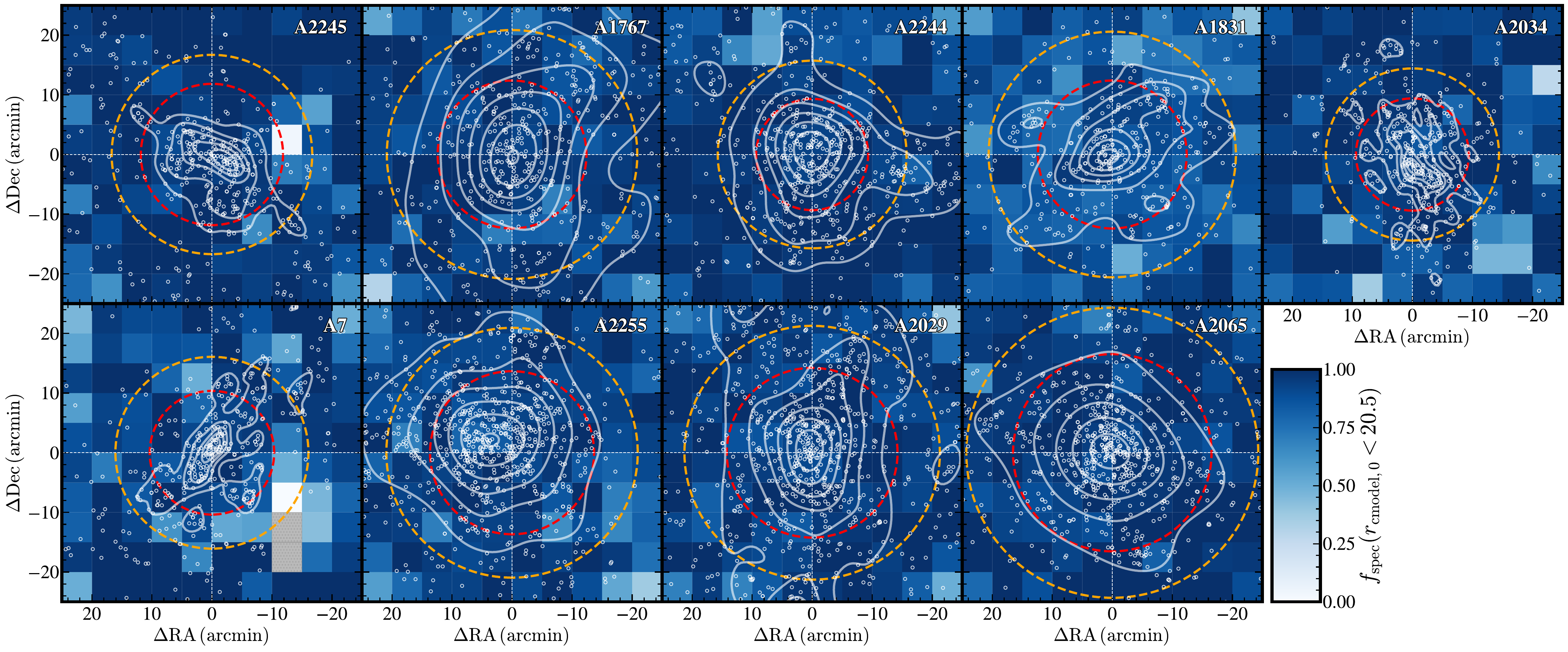} 
\caption{Spectroscopic survey completeness map for galaxies with $r_{\mathrm{cModel,0}} < 20.5$ in the MACH cluster fields. Darker colors indicate higher completeness. White points are spectroscopically identified cluster members. White contours show the projected number density of cluster members. Red and yellow dashed circles mark $R_{500}$ and $R_{200}$, respectively. }
\label{fig:2Dspec_completeness}
\end{figure*}

Figure \ref{fig:2Dspec_completeness} displays two-dimensional spectroscopic survey completeness maps for the MACH clusters. We compute the spectroscopic completeness for galaxies brighter than $r_\mathrm{cModel, 0} = 20.5$. Darker colors indicate greater completeness. Red and orange circles indicate $R_{500}$ and $R_{200}$, respectively. All of the MACH clusters are uniformly and highly complete within the survey area, particularly within $R_{200}$. For A1831, the overall survey completeness is lower, but the survey is uniform. In Figure \ref{fig:2Dspec_completeness}, we mark the distribution of spectroscopically identified members with white dots. White contours show the number density of cluster members. Section \ref{sec:cluster_membership} describes the process for cluster member identification. 

\subsubsection{Cluster Membership and Dynamical Mass} \label{sec:cluster_membership}

\centerwidetable
\begin{deluxetable*}{lccccccc}
\tablewidth{\textwidth}
\label{tab:tab2_MACHcausticinfo}
\tabletypesize{\small}
\setlength{\tabcolsep}{8pt}
\caption{The Dynamical Properties of MACH clusters and Cluster Membership}
\tablehead{
\colhead{ID} & \colhead{$R_{200}$ (Mpc)} & \colhead{$\log{(M_{200}/\Mdot)}$} & \colhead{$R_{500}$ (Mpc)} & \colhead{$\log{(M_{500}/\Mdot)}$} & \colhead{$N_{\text{mem}}$} & \colhead{$N_{\text{mem}}(<3R_{200})$} & \colhead{$N_{\text{mem}}(<R_{200})$}}
\startdata
A2245 & $1.65 \pm 0.04$ & $14.74 \pm 0.03$ & $1.17 \pm 0.03$ & $14.70 \pm 0.03$ & 548 & 539 & 312 \\
A1767 & $1.70 \pm 0.07$ & $14.78 \pm 0.03$ & $1.01 \pm 0.04$ & $14.50 \pm 0.05$ & 668 & 577 & 377 \\
A2244 & $1.72 \pm 0.07$ & $14.80 \pm 0.03$ & $1.02 \pm 0.04$ & $14.52 \pm 0.05$ & 748 & 637 & 364 \\
A1831 & $1.76 \pm 0.03$ & $14.82 \pm 0.01$ & $1.06 \pm 0.02$ & $14.56 \pm 0.02$ & 520 & 472 & 313 \\
A2034 & $1.78 \pm 0.00$ & $14.85 \pm 0.00$ & $1.16 \pm 0.01$ & $14.70 \pm 0.01$ & 410 & 410 & 287 \\
A7    & $1.83 \pm 0.04$ & $14.89 \pm 0.02$ & $1.18 \pm 0.03$ & $14.71 \pm 0.03$ & 436 & 436 & 281 \\
A2255 & $1.90 \pm 0.03$ & $14.93 \pm 0.02$ & $1.24 \pm 0.02$ & $14.76 \pm 0.02$ & 1099 & 889 & 618 \\
A2029 & $1.90 \pm 0.18$ & $14.93 \pm 0.08$ & $1.27 \pm 0.11$ & $14.80 \pm 0.11$ & 1377 & 1263 & 538 \\
A2065 & $2.03 \pm 0.08$ & $15.01 \pm 0.04$ & $1.38 \pm 0.06$ & $14.90 \pm 0.06$ & 1099 & 848 & 609 \\ \hline \hline
\enddata
\end{deluxetable*}

\begin{figure*}[htbp]
\centering
\includegraphics[width=\textwidth]{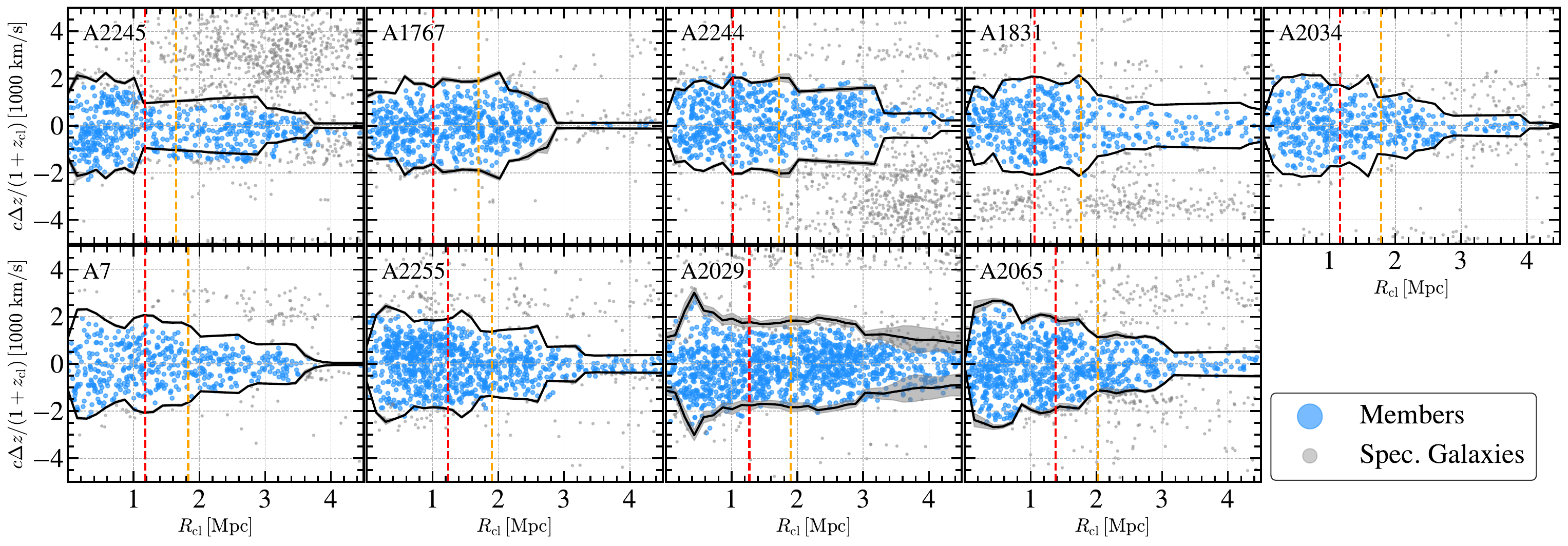}    
\caption{The R--v diagram for galaxies in the MACH cluster fields. Gray circles indicate galaxies with spectroscopic redshifts. Blue circles show member galaxies identified based on the caustic technique: black solid lines show the caustic boundaries. Red and orange vertical lines mark $R_{500}$ and $R_{200}$, respectively.} 
\label{fig:rv_dgram}
\end{figure*}

We determine cluster membership based on the caustic technique. The caustic technique computes the escape velocity profile relative to the cluster center to derive the mass profile as a function of projected distance \citep{Diaferio1997, Diaferio1999, Serra2011}. As a useful byproduct, the caustic technique determines the boundary of clusters in phase-space which enables the membership determination \citep{Diaferio1997, Diaferio1999, Serra2011, Serra2013, Kang2024}. In particular, \citet{Serra2013} investigated membership identification with the caustic technique based on  N-body simulations for clusters with $\sim 200$ observed members. They showed that the completeness of membership identification based on the caustic technique exceeds 95\% within $3R_{200}$. Many dense spectroscopic surveys use the caustic technique for cluster member identification (e.g., \citealp{Rines2006, Rines2013, Rines2016, Sohn2017, Sohn2020, Pizzardo2021, Pizzardo2024, Kang2025}).

Figure \ref{fig:rv_dgram} displays the relative rest-frame velocity as a function of  projected distance (the R-v diagram) for the MACH clusters. Solid lines in Figure \ref{fig:rv_dgram} mark the caustic boundaries. Galaxies within the caustic boundaries are the spectroscopically identified members. MACH clusters contain 410 - 1377 spectroscopically identified members with a mean of 767 members. 

The caustic technique computes the dynamical mass profile of clusters based on the escape velocity profile \citep{Diaferio1997}. \citet{Pizzardo2023a} calibrate the caustic technique using 1697 clusters with $M_{200} > 10^{14} M_{\odot}$ in the IllustrisTNG simulations. They show that the caustic technique determines the cluster mass with 10\% accuracy within $0.6 < (R_\mathrm{cl} / R_{200}) < 4.2$. Based on the caustic mass profiles, we derive the mass and radius of each MACH cluster. We first compute $R_{500}$ and $R_{200}$. We then derive the mass enclosed within these radii, $M_{500}$ and $M_{200}$. Table \ref{tab:tab2_MACHcausticinfo} summarizes the results.

\begin{figure*}[htbp]
    \centering
    \includegraphics[width=\textwidth]{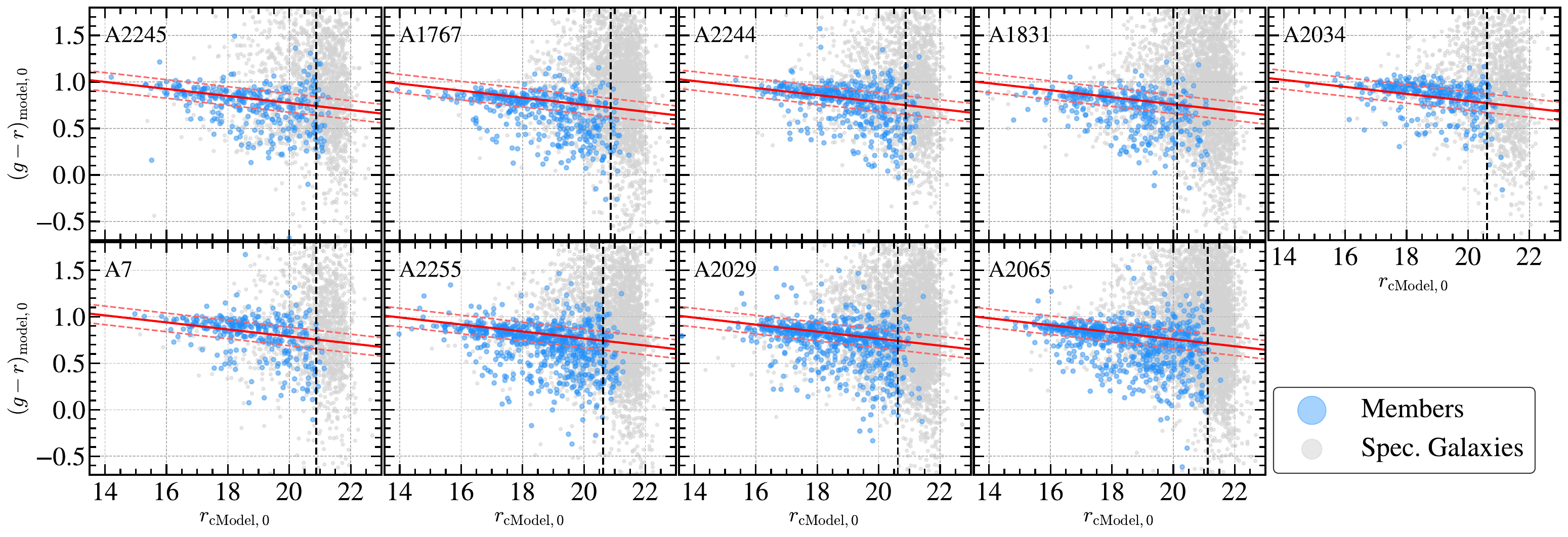}    
    \caption{$(g - r)_{0}$ vs. $r$ color–magnitude diagram for galaxies with spectroscopy in the MACH cluster fields that lie  within $R_{200}$ of each cluster center. Blue circles indicate spectroscopically identified cluster members. Gray circles represent non-members. The red solid and dashed lines show the red sequence and its boundaries, derived from the stacked sample of MACH cluster members. Black vertical lines indicate the spectroscopic survey completeness limit for each cluster.} 
    \label{fig:CMD_MACH}
\end{figure*}

Figure \ref{fig:CMD_MACH} shows the $(g-r)_{0}$ color -- $r$-band magnitude diagram for galaxies in each MACH cluster field ($R_\mathrm{cl} < R_{200}$). Blue circles indicate spectroscopically identified members; gray circles show foreground/background objects. The cluster members lie along a well-defined red sequence \citep{Visvanathan1977, Gladders2000, Rykoff2014, Rykoff2016}, with only a small fraction of blue galaxies. Importantly, the MACH spectroscopy is uniformly complete for galaxies of all colors. The survey completeness ranges from 89\% to 91\% at $r < 20.5$ for galaxies on, bluer than, or redder than the red sequence. The uniform completeness across all cluster galaxies is essential for evaluating the SMFs of diverse galaxy populations without introducing color-dependent selection biases.

We determine the characteristic red-sequence of the MACH clusters by stacking the spectroscopically confirmed member galaxies for all nine clusters to construct a $(g-r)$ versus $r$-band ($M_r$) color–magnitude diagram. To account for the different redshifts of the MACH clusters, we use the $r$-band absolute magnitudes. Following the redMaPPer cluster catalog \citep{Rykoff2016}, we use SDSS model magnitudes to compute the colors.

From the stacked sample of all MACH member galaxies with $M_{r} < -19$, we derive the best-fit linear relation:
\begin{equation} \label{eq:RedSequence}
(g - r)_{\rm model, 0} = (-0.036 \pm 0.001) M_{r} + (0.125 \pm 0.030). 
\end{equation}
The red solid lines in Figure \ref{fig:CMD_MACH} display this fit. We identify galaxies within $\Delta |(g-r)| \leq 0.1$ from the fit as red-sequence galaxies. 

We evaluate the contamination of the red sequence by superimposed foreground/background galaxies. For galaxies brighter than the spectroscopic completeness limit for each cluster, the typical contamination rate ranges from 34\% to 55\%, with a mean fraction of 45\%. In other words, about half of the photometric galaxies with a color similar to the red-sequence galaxies objects are superimposed along the line of sight. This significant contamination highlights the necessity of dense spectroscopy. 

\subsubsection[Dn4000]{$\dn$} \label{sec:dn4000}
$\dn$ is a powerful spectral indicator that traces the mean age of the stellar population of galaxies. $\dn$ is the flux ratio between two spectral ranges around $4000~\angstrom$. Young stellar populations have lower $\dn$ because multiply ionized elements in hot stars reduce the opacity thus making the $4000 ~\angstrom$ break weaker. \citet{Kauffmann2003a} show that the $\dn$ of a simple stellar population model increases monotonically when the stellar population is older than 10 Myr. Thus, the $\dn$ for a galaxy traces the mean age of underlying stellar population. 

We measure the flux ratio between $3850 - 3950~\angstrom$ and $4000$--$4100~\angstrom$ following the \citet{Balogh1999} definition. We derive $\dn$ from both SDSS and Hectospec. \citet{Fabricant2008} show that values of $\dn$ from SDSS and Hectospec spectra for duplicate target objects agree within 5\%. Because SDSS and Hectospec use fibers with different apertures (3 arcsec and 1.5 arcsec, respectively), the small difference indicates  negligible variation of $\dn$ with the physical coverage of fiber aperture. We thus apply no corrections to $\dn$ obtained from the two spectrographs.

$\dn$ is useful for separating star-forming and quiescent galaxies. Galaxies generally show a bimodal $\dn$ distribution (e.g., \citealp{Kauffmann2003a, Deshev2017, Sohn2017}); the lower $\dn$ indicates a star-forming population; the higher $\dn$ indicates a quiescent population. This classification is consistent with  other galaxy classification methods based on $\rm{H\alpha}$ emission line and the UVJ color--color diagrams (e.g., \citealp{Kauffmann2003a, Woods2010, Damjanov2015}). Following \citet{Sohn2017} and \citet{Sohn2019b}, we define quiescent galaxies as objects with $\dn > 1.5$ in all of the MACH clusters.  

\subsubsection{Stellar Mass} \label{sec:stellar_mass}

\begin{deluxetable*}{lcr}
\tablewidth{\textwidth}
\label{tab:cigale_params}
\setlength{\tabcolsep}{15pt}
\renewcommand{\arraystretch}{0.9}
\caption{CIGALE SED Fitting Parameters}
\tablehead{
\colhead{Parameter} & \colhead{Symbol} & \colhead{Values}
}
\startdata
\multicolumn{3}{c}{Stellar population models: \citep{Bruzual2003}} \\
\hline
Initial mass function & IMF & \citet{Chabrier2003} \\
Metallicity & $Z$ & 0.0001, 0.0004, 0.004, 0.008, 0.02, 0.05 \\
Separation age & $t_{\mathrm{sep}}$ & 10 Myr \\
\hline
\multicolumn{3}{c}{SFH: Delayed with optional burst} \\
\hline
e-folding time of main population & $\tau_{\mathrm{main}}$ & 10, 74, 547, 4053, 30000 Myr \\
Age of main population & $t_{\mathrm{main}}$ & 10 - 13000 Myr (30 logarithmic bins) \\
e-folding time of burst & $\tau_{\mathrm{burst}}$ & 50.0 Myr \\
Age of burst & $t_{\mathrm{burst}}$ & 10 Myr \\
Mass fraction of burst & $f_{\mathrm{burst}}$ & 0.0 \\
\hline
\multicolumn{3}{c}{Dust attenuation: Modified Starburst \citep{Calzetti2000}} \\
\hline
Color excess of nebular lines & $E(B-V)_{\mathrm{lines}}$ & 0.0, 0.05, 0.1, 0.2, 0.3, 0.4, 0.6 \\
Reduction factor & $E(B-V)_{\mathrm{star}} / E(B-V)_{\mathrm{lines}}$ & 0.44 \\
UV bump amplitude & $B_{\mathrm{amp}}$ & 0.0 \\
Power law slope & $\delta$ & 0.0 \\
$R_V=A_V/E(B-V)$ & $R_V$ & 3.1 \\ \hline \hline
\enddata
\end{deluxetable*}

We derive galaxy stellar masses based on CIGALE (Code Investigating GALaxy Emission, \citealp{Boquien2019}). CIGALE is a spectral energy distribution (SED) fitting code based on observed fluxes. We use SDSS $ugriz$ photometry and spectroscopic redshifts for MACH cluster galaxies to estimate the stellar masses. 

CIGALE generates SED models based on various star formation histories (SFH), stellar population synthesis (SPS) models, and dust attenuation laws. We use models similar to those used for studying the A2029 galaxy population (i.e., \citealp{Sohn2017}). We use a delayed star formation model with an optional exponential burst for the star formation history. We assume an $e-$folding time ranging from 0.01 to 30 Gyr. The age of the main stellar populations ranges from 0.01 to 13 Gyr. Both the star formation rate and  the age  are binned logarithmically. We also use the SPS model from \citet{Bruzual2003} with an initial mass function (IMF) from \citet{Chabrier2003}. We assume six metallicities for the SPS models: 0.0001, 0.0004, 0.004, 0.008, 0.02, and 0.05. We use the extinction law from \citet{Calzetti2000}. We select $E(B-V)$ values ranging from 0.0 to 0.6, divided into seven bins. Table \ref{tab:cigale_params} summarizes the SED fitting parameters.

We derive the best-fit models using a likelihood-weighted mean (Bayesian approach) in CIGALE. CIGALE first computes $\chi^2$ by comparing the observed photometry with the SED models. CIGALE converts the $\chi^{2}$ into probabilities using $P_{i} = \exp\left(-{\chi_i^2}/{2}\right)$. Models with lower $\chi^{2}$ have higher $P_{i}$ values. Based on the $P_{i}$ distributions computed from all of the SED model comparisons, CIGALE determines the likelihood-weighted mean of the parameters, including stellar mass, metallicity, and stellar population age. We use \texttt{bayes.stellar.m\_star} as the stellar mass estimate. The stellar mass uncertainty is the $1\sigma$ standard deviation of stellar masses in the likelihood-weighted distributions. 

We compare the CIGALE stellar mass estimates with those from the MPA/JHU catalog \citep{Kauffmann2003a, Salim2007}. For this comparison, we use the CIGALE stellar mass estimates for 170230 galaxies in the SDSS Main Galaxy Sample (MGS) in the redshift range $0.07 < z < 0.11$ (see Section \ref{sec:SDSS_MGS}). There are 165538 galaxies with stellar mass estimates from the MPA/JHU catalog. Hereafter, we refer to the stellar mass from the MPA/JHU catalog as the MPA stellar mass.

\begin{figure}[htbp]
\centering
\includegraphics[width=0.48\textwidth]{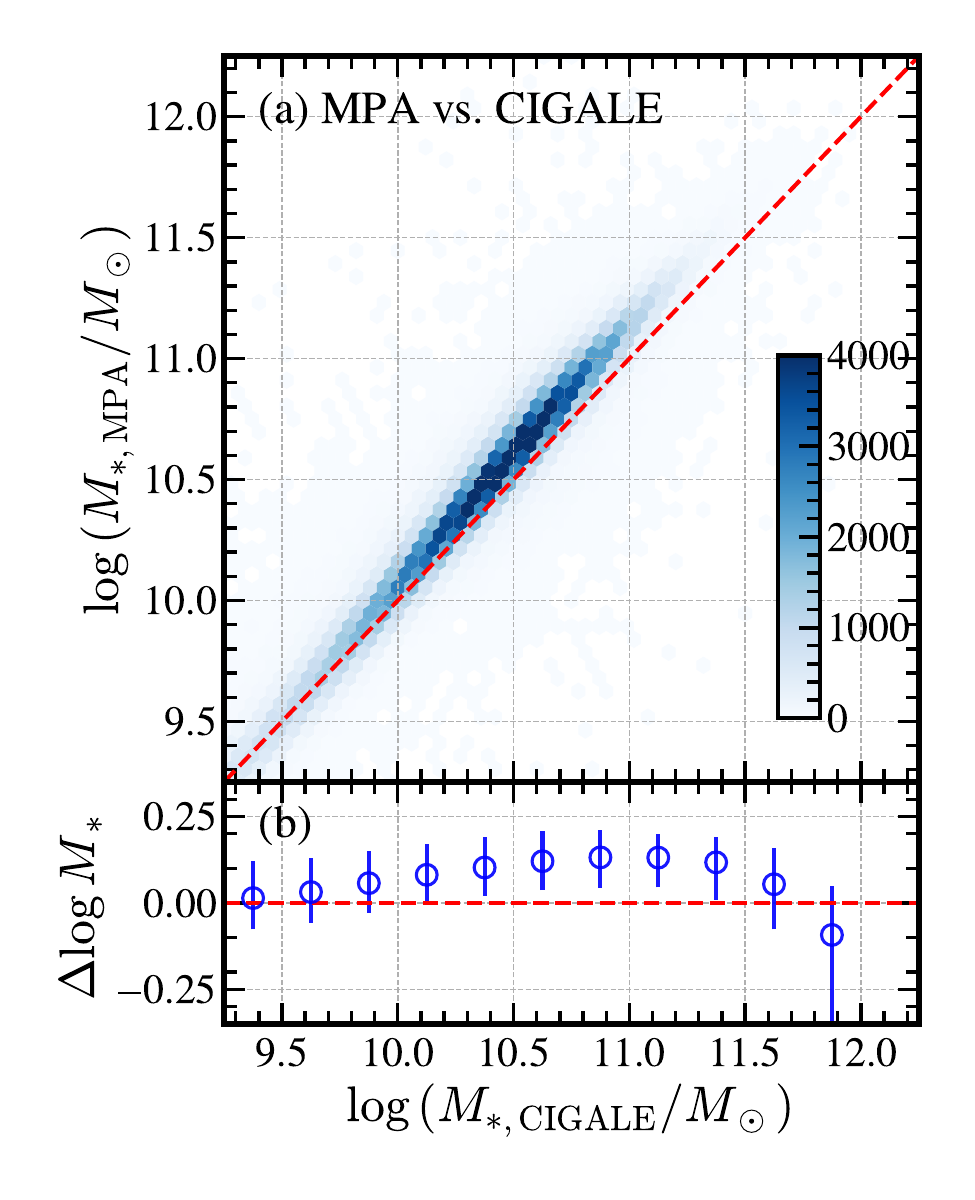}
\caption{(a) Comparison between stellar masses from the MPA/JHU catalog and from CIGALE (i.e., this study). Darker colors indicate higher galaxy densities. The red dashed line marks the one-to-one relation. (b) Median offset between MPA and CIGALE stellar masses. Error bars indicate the interquartile range (Q1–Q3).}  
\label{fig:Mstellar_tool_offset}
\end{figure}

Figure \ref{fig:Mstellar_tool_offset} compares the CIGALE and MPA stellar masses (blue contours). The MPA stellar masses are generally 0.11 dex larger than the CIGALE stellar masses for $\logMdot<11.4$. For $\logMdot > 11.5$, the MPA stellar masses lie slightly below the CIGALE stellar masses, but the number of galaxies is small ($\sim 150$ galaxies). The systematic offset is comparable with the uncertainty in individual stellar mass estimates; the typical uncertainty in MPA and CIGALE stellar mass estimates are 0.10 dex and 0.14 dex, respectively. 

\begin{figure*}[t]
\centering
\includegraphics[width=0.8\textwidth]{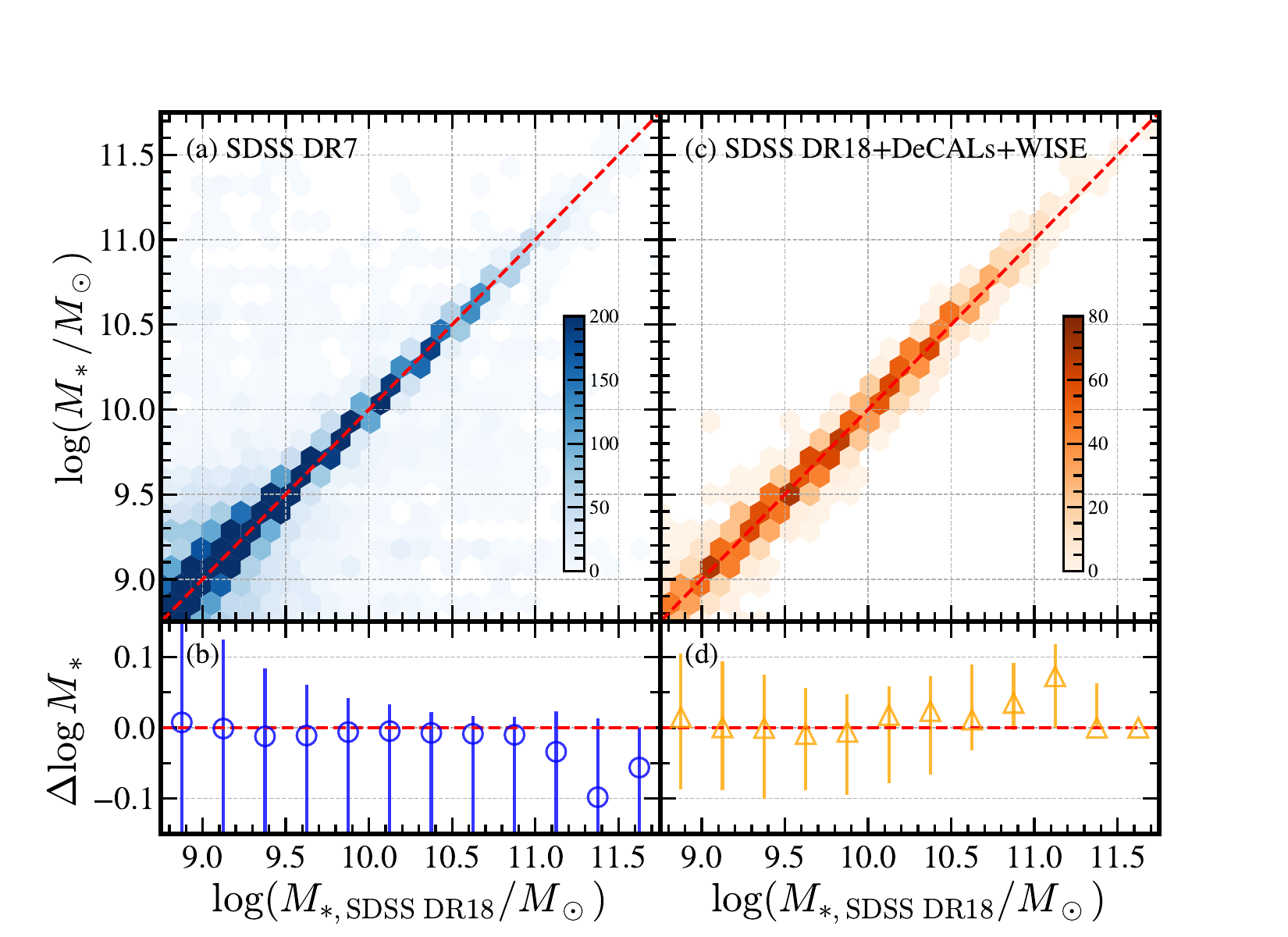}
\caption{(a) Comparison between stellar mass estimates for galaxies in the MACH cluster fields based on SDSS DR7 and DR18 photometry. The bluer contours indicate higher number density. The red dashed line shows the one-to-one relation. (b) Median stellar mass offset ($\rm M_{*, \rm{SDSS, ~DR7}} - M_{*, \rm{SDSS,~DR18}}$) as a function of SDSS DR18 stellar mass. Error bars show the interquartile range (Q1–Q3). (c), (d) Same as the (a), (b), respectively, but for a comparison between stellar mass estimates based on SDSS $ugriz$ and (SDSS $ugriz$ + DECaLS $grz$ + WISE) photometry. } 
\label{fig:Mstellar_photband_offset}
\end{figure*}

We further test the CIGALE stellar masses based on DR7 and DR18 photometry. Because MPA is based on an earlier version of SDSS photometry, this test enables identification of the origin of the difference between the CIGALE stellar masses we use and the MPA stellar masses. 

Figure \ref{fig:Mstellar_photband_offset} (a) and (b) compare stellar mass estimates based on SDSS DR18 and DR7 photometry. The mass estimates based on the two versions of SDSS photometry are generally similar. The mass differences in Figure \ref{fig:Mstellar_photband_offset} are consistent with zero over a wide mass range. This comparison demonstrates that the difference between the CIGALE and MPA stellar mass estimates does not result from the difference in photometry; the SED fitting techniques and the models used for stellar mass estimates are responsible for the mass differences. 

We also examine the stellar mass estimates based on more extended photometric data. For MACH cluster galaxies, we compile $grz$ photometry from the DECaLS survey \citep{Dey2019} and photometry from the Wide-Field Infrared Survey Explorer (WISE; \citealp{Wright2010}) in the $W_1$ (3.4 $\mu$m) and $W_2$ (4.6 $\mu$m) bands. We then derive stellar masses with the same setup. Figure~\ref{fig:Mstellar_photband_offset} (c) and (d) compare stellar mass estimates based on DR18 photometry only with those based on photometry including DR18, DECaLS, and WISE. We find no systematic differences when  we use the extended photometric data. These comparisons confirm that the CIGALE stellar mass estimates remain robust regardless of the choice of photometry.

\subsection{SDSS Field Comparison Sample} \label{sec:SDSS_MGS}

We next build an SDSS field galaxy sample to  compare the field and cluster SMFs. We use the SDSS Main Galaxy Sample (MGS, \citealp{Strauss2002}). We select photometric galaxies with \texttt{probPSF} = 0 and remove saturated sources. We also perform visual inspections for 62972 objects with $r < 17.77$ to remove stars, spikes, bleeding, and fragments of bright objects. We removed 50761 objects with bad photometry. 

We then select MGS galaxies within the redshift range of the MACH clusters (i.e., $0.07 < z < 0.11$). We select MGS galaxies in the Northern Galactic Pole (NGP) region. The NGP region covers 8032 deg$^{2}$. For the NGP region, we obtain both photometry and spectroscopy from SDSS DR18. The SDSS spectroscopy is 92.3\% complete for galaxies brighter than $r_{\rm petro} = 17.77$ and with half-light surface brightness ($\mu_{50}$) brighter than 24.5 mag arcsec$^{-2}$. Thus, SDSS MGS does not require significant correction for incompleteness when we measure the SMF.

In total, the SDSS field galaxy sample includes 170225 galaxies with $0.07 < z < 0.11$. We measure the stellar mass (see Section \ref{sec:stellar_mass}) and $\dn$ of these SDSS field galaxies with the same procedure as for the MACH cluster galaxies. The stellar mass and $\dn$ measurements for SDSS MGS/NGP galaxies are 99.1\% and 99.9\% complete, respectively. Hereafter, we refer to this sample as the SDSS field sample. 

\section{The Cluster SMF} \label{sec:cluster_smf}

The stellar mass function (SMF) is the number density of galaxies as a function of stellar mass. We derive the cluster SMFs based on the set of spectroscopically identified member galaxies in MACH clusters. We use CIGALE stellar mass estimates. The cluster SMFs may be affected by incompleteness. Thus in Section \ref{sec:comp_cor_smf}, we derive cluster SMFs and correct for the MACH spectroscopic survey incompleteness. In Section \ref{sec:mach_cluster_smf}, we display the MACH cluster SMFs.  

\subsection{SMF Derivation} \label{sec:comp_cor_smf}

MACH is a dense and complete spectroscopic survey of local massive galaxy clusters. Of course the MACH spectroscopy is not 100\% complete. Furthermore, the incompleteness of the spectroscopic survey as a function of magnitude does not translate directly into the stellar mass completeness. 

\begin{figure*}[htbp]
\centering
\includegraphics[width=1.0\textwidth]{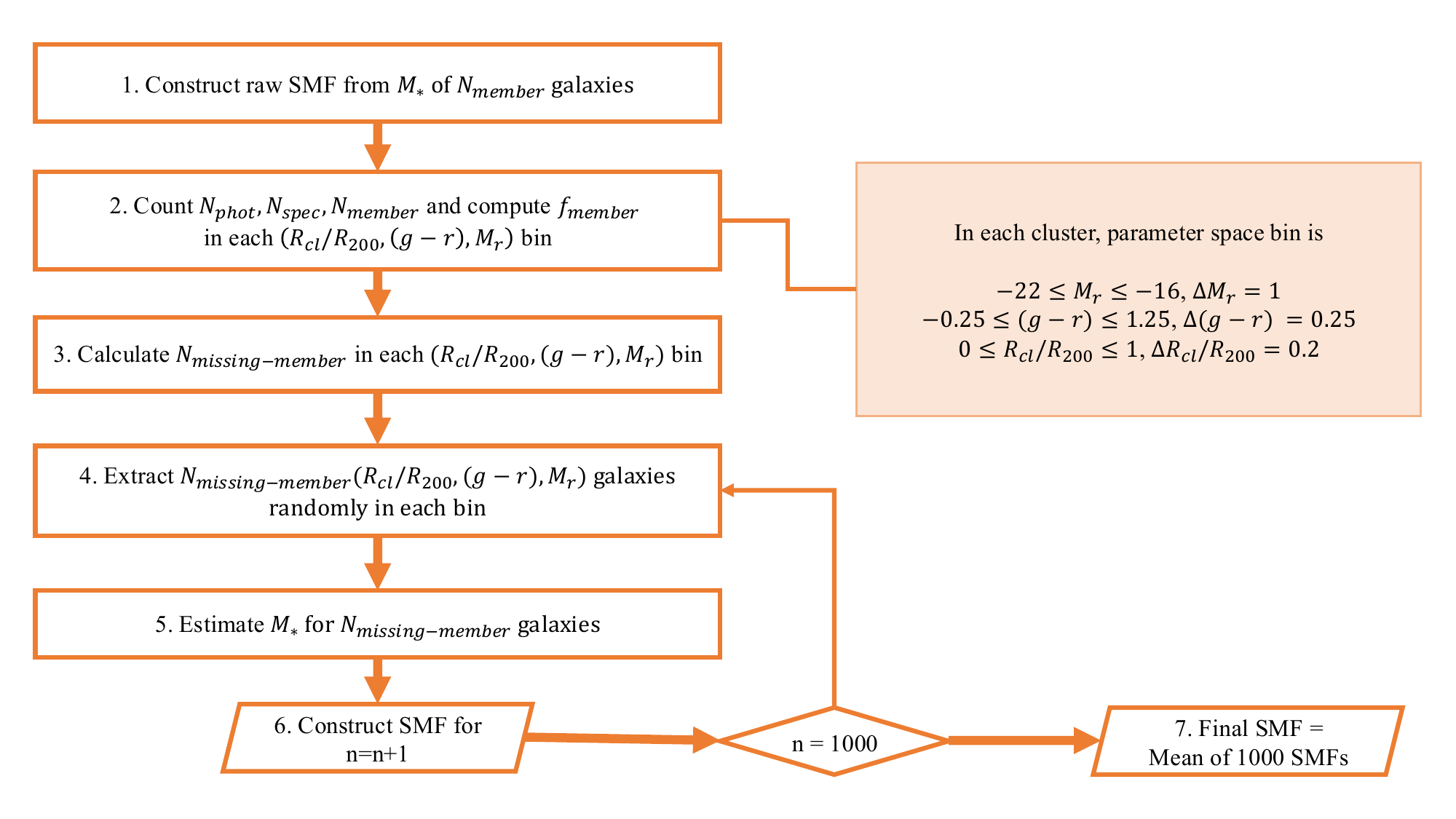}
\caption{Flowchart summarizing the SMF construction process.}
\label{fig:FlowChart}
\end{figure*}

We derive MACH SMFs by applying a correction for spectroscopic incompleteness. Figure \ref{fig:FlowChart} summarizes our procedure  for deriving the MACH SMF including the incompleteness correction. The correction procedure includes 1) estimating the number of cluster members in the cluster fields without spectroscopy, 2) assigning a membership probability for these photometric galaxies without spectroscopy, and 3) measuring the stellar mass of the potential photometric members. We discuss the details below. 

We first construct the raw SMF based on galaxies within $R_\mathrm{cl} < R_{200}$ (step 1 in Figure \ref{fig:FlowChart}). For these galaxies, we also derive various physical properties including $(g-r)$ color and $r$-band absolute magnitude ($M_{r}$). There are photometric galaxies without spectroscopy within $R_{200}$. Some of these galaxies may be cluster members, but we do not know for sure due to the lack of spectroscopy. We compute the membership probability based on the MACH stacked sample, including all galaxies within $R_\mathrm{cl} < R_{200}$ in the MACH fields. The probability of being a member of a cluster varies depending on $R_\mathrm{cl}$, $(g-r)$, and $M_{r}$. To compute this probability, we use all galaxies within $R_{200}$ in the MACH cluster samples. We divide the stacked sample into subsamples within various $M_{r}$, $(g-r)$, and $R_\mathrm{cl}$ bins. We then count the number of photometric galaxies ($N_\mathrm{phot}$), spectroscopic galaxies ($N_\mathrm{spec}$), and spectroscopically identified member galaxies ($N_\mathrm{mem}$) in each subsample (step 2 in Figure \ref{fig:FlowChart}).

\begin{figure*}[htbp]
\centering
\includegraphics[width=\textwidth]{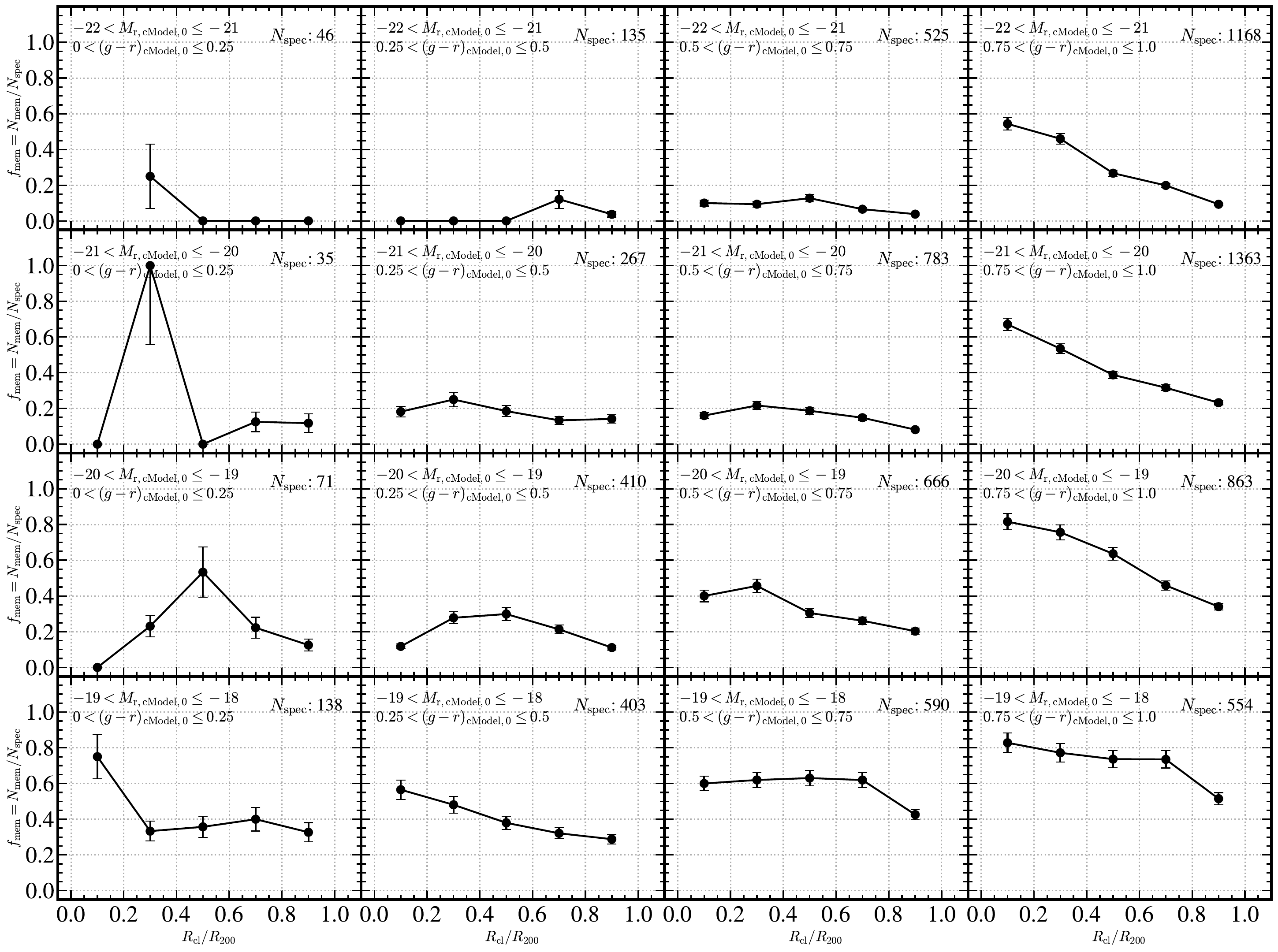}
\caption{The fraction of member galaxies among galaxies with spectroscopy in each absolute magnitude and color bin as a function of normalized cluster-centric distance based on the stacked spectroscopic sample for all of the MACH clusters. The number of spectroscopic objects in each bin is indicated in each panel. Error bars represent Poisson errors.} 
\label{fig:SMF_correction_params}
\end{figure*}

Figure \ref{fig:SMF_correction_params} shows the member fraction among all MACH spectroscopic galaxies in subsamples with various $M_{r}$ and $(g-r)$ ranges as a function of $R_\mathrm{cl}$. We use cluster-centric distance normalized by $R_{200}$ as a dimensionless unit, $R' = R_\mathrm{cl}/R_{200}$. The member fraction is:
\begin{equation}
\label{eq:fmem}
f_\mathrm{mem} (M_{r}, (g-r), R') = \frac{N_\mathrm{mem} (M_{r}, (g-r), R')}{N_\mathrm{spec} (M_{r}, (g-r), R')}
\end{equation}

The membership fraction varies significantly with $M_{r}$, $(g-r)$, and $R_\mathrm{cl}$. Thus this differential correction is critical. Within the same $M_{r}$ and $(g-r)$ bins, the membership fraction generally decreases toward larger $R_\mathrm{cl}$. The membership fractions are significantly higher in the $0.75 < (g-r) < 1.0$ bins where the red-sequence of the MACH clusters lies. 

Based on the membership probability function, we determine the number of missing members ($N_\mathrm{missing\text{-}member}$) in each bin (step 3 in Figure \ref{fig:FlowChart}):
\begin{equation}
    N_\mathrm{missing\text{-}member} = (N_\mathrm{phot} - N_\mathrm{spec}) \times f_\mathrm{mem}
\end{equation}
To determine $N_\mathrm{missing\text{-}member}$, we calculate $M_{r}$ for photometric galaxies without spectroscopy by assuming that they are all at the cluster redshift. 

This approach shows that $N_\mathrm{missing\text{-}member}$ galaxies should be added in order to complete the sample. Because we do not know which photometric galaxies are the missing member galaxies, we randomly select $N_\mathrm{missing\text{-}member}$ photometric galaxies without spectroscopy in bins of $M_{r}, (g-r)$, and $R'$ bins (step 4 in Figure \ref{fig:FlowChart}). After randomly selecting these members, we measure their stellar masses using the procedure applied to the spectroscopic members (step 5 in Figure \ref{fig:FlowChart}).

The final SMF is based on the sample that includes both the spectroscopically identified members and the possible member galaxies we selected from the photometric sample. We repeat this identification of probable photometric members 1000 times. The mean of the 1000 corrected SMFs is the final SMF of the cluster (steps 6 and 7 in Figure \ref{fig:FlowChart}).

\subsection{MACH SMF} \label{sec:mach_cluster_smf}

\begin{figure*}[htbp]
    \centering
    \includegraphics[width=\textwidth]{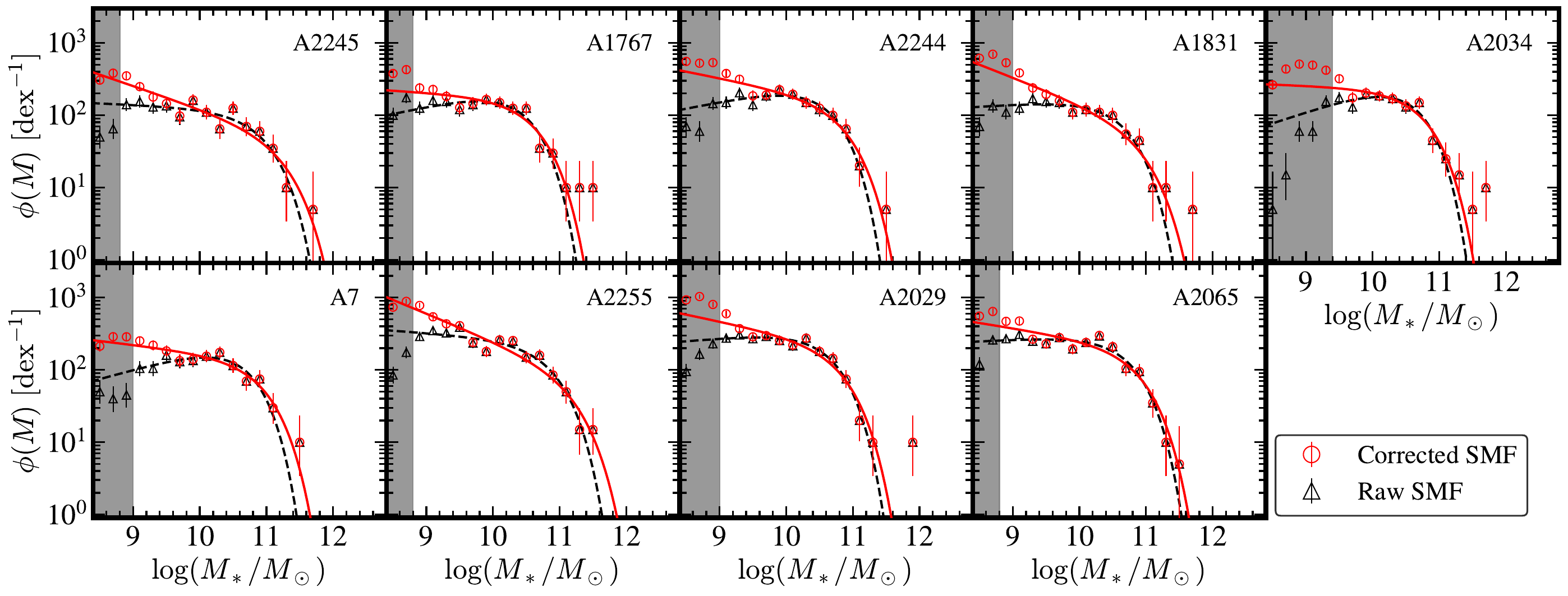}
    \caption{SMF of the MACH clusters. Black triangles and dashed lines show the raw SMF and its best-fit Schechter function. Red squares and solid lines show the SMF corrected for spectroscopic incompleteness and its best-fit Schechter function. Gray-shaded regions indicate stellar mass ranges where the incompleteness correction is larger than 10\%.}
    \label{fig:MACH_SMF}
\end{figure*}

\begin{figure}[htbp]
    \centering
    \includegraphics[width=0.5\textwidth]{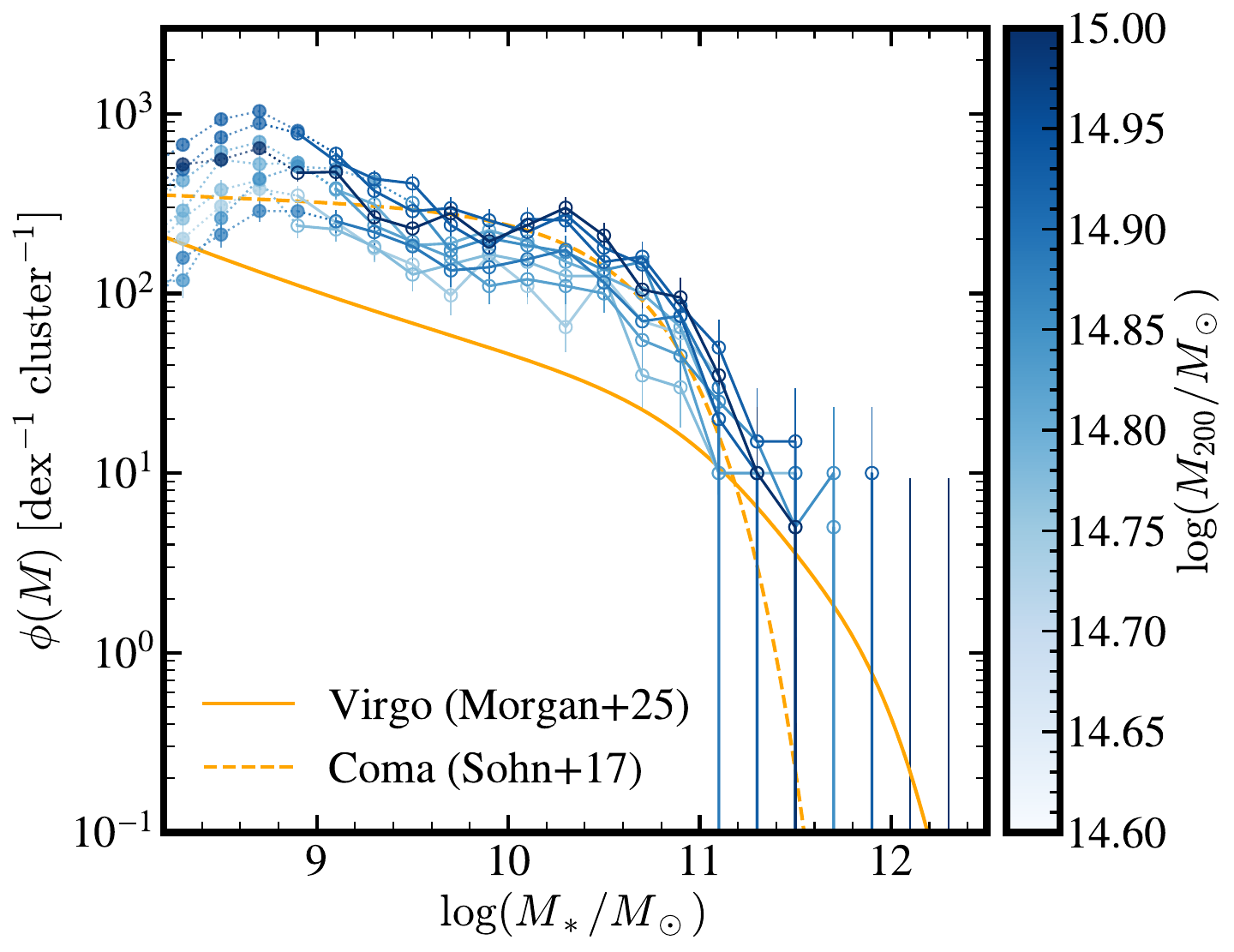}
    \caption{SMFs of the MACH clusters (open circles and solid lines) after the incompleteness correction; filled circles (dotted lines) indicate the mass range with lower completeness. For comparison, the Virgo \citep{Morgan2025} and Coma \citep{Sohn2017} SMFs are shown as orange solid and dashed lines, respectively.}
    \label{fig:corSMF_onepanel}
\end{figure}

Figure \ref{fig:MACH_SMF} displays the raw and corrected SMFs of the MACH clusters. The raw and corrected SMFs are consistent with each other for $\logMdot > 9.5$. The completeness correction becomes significant for $\logMdot \lesssim 9$. We indicate the stellar mass ranges where the SMF determination is based on incomplete spectroscopy  with gray shading. At these low stellar masses, the incompleteness correction exceeds 10\% of the raw SMF. In other words, the completeness correction affects the determination of the shape of the SMF significantly in the shaded region. Table \ref{tab:tab_SMF_data} lists the raw and corrected SMFs for four example MACH clusters; the SMFs for the remaining MACH clusters are available online. 

\begin{deluxetable*}{ccccccccc}
\tablewidth{\textwidth}
\label{tab:tab_SMF_data}
\setlength{\tabcolsep}{3.5pt}
\renewcommand{\arraystretch}{0.9}
\caption{SMFs for MACH Clusters$^{*}$}
\tablehead{
\multirow{2}{*}{$\logMdot$} &
\multicolumn{2}{c}{A2245} &
\multicolumn{2}{c}{A1767} &
\multicolumn{2}{c}{A2244} &
\multicolumn{2}{c}{A1831} \\
& $\phi_{\mathrm{raw}}$ & $\phi_{\mathrm{cor}}$
& $\phi_{\mathrm{raw}}$ & $\phi_{\mathrm{cor}}$
& $\phi_{\mathrm{raw}}$ & $\phi_{\mathrm{cor}}$
& $\phi_{\mathrm{raw}}$ & $\phi_{\mathrm{cor}}$
}
\startdata
$8.5$  & $50.0 \pm 15.8$ & $305.9 \pm 39.1$ & $100.0 \pm 22.4$ & $377.6 \pm 43.5$ & $70.0 \pm 18.7$ & $554.8 \pm 52.7$ & $70.0 \pm 18.7$ & $613.1 \pm 55.4$ \\
$8.7$  & $65.0 \pm 18.0$ & $381.4 \pm 43.7$ & $175.0 \pm 29.6$ & $426.8 \pm 46.2$ & $60.0 \pm 17.3$ & $524.3 \pm 51.2$ & $135.0 \pm 26.0$ & $697.1 \pm 59.0$ \\
$8.9$  & $140.0 \pm 26.5$ & $350.1 \pm 41.8$ & $125.0 \pm 25.0$ & $238.1 \pm 34.5$ & $145.0 \pm 26.9$ & $534.4 \pm 51.7$ & $110.0 \pm 23.5$ & $531.5 \pm 51.5$ \\
$9.1$  & $155.0 \pm 27.8$ & $248.3 \pm 35.2$ & $160.0 \pm 28.3$ & $227.2 \pm 33.7$ & $150.0 \pm 27.4$ & $377.8 \pm 43.5$ & $125.0 \pm 25.0$ & $384.5 \pm 43.8$ \\
$9.3$  & $130.0 \pm 25.5$ & $178.3 \pm 29.9$ & $155.0 \pm 27.8$ & $182.1 \pm 30.2$ & $205.0 \pm 32.0$ & $314.0 \pm 39.6$ & $170.0 \pm 29.2$ & $238.5 \pm 34.5$ \\
$9.5$  & $135.0 \pm 26.0$ & $145.1 \pm 26.9$ & $120.0 \pm 24.5$ & $127.4 \pm 25.2$ & $140.0 \pm 26.5$ & $185.2 \pm 30.4$ & $155.0 \pm 27.8$ & $193.9 \pm 31.1$ \\
$9.7$  & $95.0 \pm 21.8$ & $97.6 \pm 22.1$ & $140.0 \pm 26.5$ & $145.0 \pm 26.9$ & $185.0 \pm 30.4$ & $190.0 \pm 30.8$ & $150.0 \pm 27.4$ & $158.4 \pm 28.1$ \\
$9.9$  & $160.0 \pm 28.3$ & $162.4 \pm 28.5$ & $165.0 \pm 28.7$ & $165.0 \pm 28.7$ & $225.0 \pm 33.5$ & $225.0 \pm 33.5$ & $110.0 \pm 23.5$ & $110.0 \pm 23.5$ \\
$10.1$ & $110.0 \pm 23.5$ & $110.0 \pm 23.5$ & $150.0 \pm 27.4$ & $150.0 \pm 27.4$ & $195.0 \pm 31.2$ & $195.0 \pm 31.2$ & $120.0 \pm 24.5$ & $120.0 \pm 24.5$ \\
$10.3$ & $65.0 \pm 18.0$ & $65.0 \pm 18.0$ & $125.0 \pm 25.0$ & $125.0 \pm 25.0$ & $150.0 \pm 27.4$ & $150.0 \pm 27.4$ & $110.0 \pm 23.5$ & $110.0 \pm 23.5$ \\
$10.5$ & $125.0 \pm 25.0$ & $125.0 \pm 25.0$ & $125.0 \pm 25.0$ & $125.0 \pm 25.0$ & $120.0 \pm 24.5$ & $125.0 \pm 25.0$ & $100.0 \pm 22.4$ & $100.0 \pm 22.4$ \\
$10.7$ & $70.0 \pm 18.7$ & $70.0 \pm 18.7$ & $35.0 \pm 13.2$ & $35.0 \pm 13.2$ & $100.0 \pm 22.4$ & $100.0 \pm 22.4$ & $55.0 \pm 16.6$ & $55.0 \pm 16.6$ \\
$10.9$ & $60.0 \pm 17.3$ & $60.0 \pm 17.3$ & $30.0 \pm 12.2$ & $30.0 \pm 12.2$ & $65.0 \pm 18.0$ & $65.0 \pm 18.0$ & $45.0 \pm 15.0$ & $45.0 \pm 15.0$ \\
$11.1$ & $35.0 \pm 13.2$ & $35.0 \pm 13.2$ & $10.0 \pm 7.1$ & $10.0 \pm 7.1$ & $20.0 \pm 10.0$ & $20.0 \pm 10.0$ & $10.0 \pm 7.1$ & $10.0 \pm 7.1$ \\
$11.3$ & $10.0 \pm 7.1$ & $10.0 \pm 7.1$ & $10.0 \pm 7.1$ & $10.0 \pm 7.1$ & - & - & $10.0 \pm 7.1$ & $10.0 \pm 7.1$ \\
$11.5$ & - & - & $10.0 \pm 7.1$ & $10.0 \pm 7.1$ & $5.0 \pm 5.0$ & $5.0 \pm 5.0$ & - & - \\
$11.7$ & $5.0 \pm 5.0$ & $5.0 \pm 5.0$ & - & - & - & - & $5.0 \pm 5.0$ & $5.0 \pm 5.0$ \\
$11.9$ & - & - & - & - & - & - & - & - \\
$12.1$ & - & - & - & - & - & - & - & - \\
$12.3$ & - & - & - & - & - & - & - & - \\
$12.5$ & - & - & - & - & - & - & - & - \\ \hline \hline
\enddata
\tablenotetext{*}{The SMFs for the remaining five MACH clusters are available online.}
\end{deluxetable*}

Above the completeness limit, we derive the best-fit Schechter function \citep{Schechter1976} to the shape of the SMF for each cluster. The Schechter function is:
\begin{equation}
\label{eq:schechter}
\phi(M) = \phi_0 \left( \frac{M^*}{M} \right)^{1+\alpha} 
\exp\left( -\frac{M^*}{M} \right), 
\end{equation}
where $\alpha$ is the slope of the power law at low stellar mass, and $M^{*}$ is the characteristic mass where the Schechter function slope changes. $\phi_{0}$ is a normalization factor. To derive the SMF parameters, we counted the number of galaxies in stellar mass bins with a width $\Delta \log(M_\ast/M_\odot) = 0.2$. We then fitted the Schechter function to the binned SMF based on a Markov Chain Monte Carlo (MCMC) approach, incorporating Poisson errors for each bin, using \texttt{emcee} package \citep{ForemanMackey2013}. The priors we adopted for the fit are $\phi_{0} \in (0,10^{4})$, $\alpha \in (-5,5)$, and $M^{*} \in (10^8,10^{14})\Mdot$. At each step of the MCMC analysis, we evaluate the Gaussian log-likelihood. Table \ref{tab:tab_SMFparams} summarizes the best-fit parameters for the raw and corrected SMFs.

In Figure \ref{fig:corSMF_onepanel}, we compare the MACH SMFs with the SMFs of Coma \citep{Sohn2017} and Virgo \citep{Morgan2025}. \citet{Sohn2017} derive the Coma SMF within $R_{200}$ based on SDSS photometry and spectroscopy. We compute the Coma SMF at $\logMdot > 9$ based on the best-fit Schechter function parameters listed in \citet{Sohn2017}. We also compute the Virgo SMF within $R_{200}$ based on the Schechter function fit parameters in Table 3 in \citet{Morgan2025}. Because \citet{Morgan2025} measured the SMF at three projected radial bins (i.e., $R < 0.5$ Mpc and $0.5 < R < 1.0$ Mpc, and $1.0 < R < 1.5$ Mpc), we combine the SMFs within $R < 1.0$ Mpc, corresponding to Virgo $R_{200}$ \citep{Simionescu2017}.

For $\logMdot \geq 11.2$, the cluster SMFs exhibit large scatter due to variations in the stellar masses of the BCGs. The SMF slopes also vary among clusters for $\logMdot < 10$, primarily reflecting differences in the relative contributions of star-forming galaxies within each cluster sample. For example, the Virgo SMF is much steeper than the Coma cluster SMF at the low-mass end, due to the higher fraction of star-forming galaxies in the Virgo cluster. Similarly, MACH clusters with a larger star-forming fractions tend to show steeper SMFs at low mass. We further discuss the contributions of star-forming and quiescent populations to the cluster SMFs in Section \ref{sec:populations}. We also note that the amplitude of Virgo SMF is significantly smaller than the other clusters, resulting from the smaller total mass of Virgo.

\begin{deluxetable*}{cccccccc}
    \tablewidth{\textwidth}
    \label{tab:tab_SMFparams}
    \setlength{\tabcolsep}{8pt}
    \caption{Best-fit Schechter Function Parameters for the MACH SMFs.}
    \tablehead{
    \multirow{2}{*}{Cluster} &
    \multicolumn{3}{c}{Raw SMF} &
    \multicolumn{3}{c}{Corrected SMF} &
    \multirow{2}{*}{$\log (\rm M_{*, \rm{limit}} /\Mdot)$\tablenotemark{$\dagger$}} \\
    \cline{2-4} \cline{5-7}
    \colhead{} &
    \colhead{$\phi_0$} & \colhead{$\logcharMass$} & \colhead{$\alpha$} &
    \colhead{$\phi_0$} & \colhead{$\logcharMass$} & \colhead{$\alpha$} &
    \colhead{}
    }
    \startdata
        A2245 & $50.13^{+19.44}_{-15.59}$ & $10.98^{+0.21}_{-0.19}$ & $-1.04^{+0.10}_{-0.09}$ & $20.70^{+24.22}_{-10.04}$ & $11.31^{+0.42}_{-0.40}$ & $-1.31^{+0.13}_{-0.08}$ & $8.8$ \\
        A1767 & $113.99^{+30.44}_{-27.81}$ & $10.47^{+0.14}_{-0.13}$ & $-0.80^{+0.12}_{-0.12}$ & $72.10^{+45.53}_{-24.99}$ & $10.66^{+0.21}_{-0.22}$ & $-1.06^{+0.17}_{-0.12}$ & $8.8$ \\
        A2244 & $132.87^{+32.99}_{-29.27}$ & $10.62^{+0.13}_{-0.13}$ & $-0.82^{+0.11}_{-0.10}$ & $63.67^{+31.05}_{-22.35}$ & $10.91^{+0.21}_{-0.19}$ & $-1.18^{+0.11}_{-0.10}$ & $9.0$ \\
        A1831 & $76.96^{+25.84}_{-22.71}$ & $10.67^{+0.17}_{-0.16}$ & $-0.94^{+0.13}_{-0.12}$ & $27.50^{+26.98}_{-13.90}$ & $11.01^{+0.35}_{-0.34}$ & $-1.36^{+0.14}_{-0.10}$ & $9.0$ \\
        A2034 & $152.30^{+40.45}_{-39.97}$ & $10.60^{+0.15}_{-0.13}$ & $-0.69^{+0.19}_{-0.17}$ & $97.25^{+49.82}_{-36.19}$ & $10.78^{+0.20}_{-0.18}$ & $-1.03^{+0.20}_{-0.16}$ & $9.4$ \\
        A7 & $118.23^{+31.68}_{-29.00}$ & $10.67^{+0.17}_{-0.15}$ & $-0.74^{+0.14}_{-0.12}$ & $58.72^{+46.82}_{-22.77}$ & $10.98^{+0.30}_{-0.29}$ & $-1.11^{+0.18}_{-0.11}$ & $9.0$ \\
        A2255 & $104.86^{+23.40}_{-20.70}$ & $10.90^{+0.11}_{-0.11}$ & $-1.07^{+0.06}_{-0.05}$ & $36.06^{+25.01}_{-13.32}$ & $11.26^{+0.24}_{-0.26}$ & $-1.38^{+0.08}_{-0.06}$ & $8.8$ \\
        A2029 & $160.58^{+34.57}_{-30.15}$ & $10.68^{+0.10}_{-0.09}$ & $-0.92^{+0.08}_{-0.08}$ & $89.64^{+31.95}_{-24.59}$ & $10.87^{+0.14}_{-0.13}$ & $-1.19^{+0.09}_{-0.08}$ & $9.0$ \\
        A2065 & $139.37^{+27.92}_{-25.31}$ & $10.76^{+0.10}_{-0.10}$ & $-0.95^{+0.07}_{-0.06}$ & $85.84^{+28.29}_{-20.58}$ & $10.93^{+0.12}_{-0.13}$ & $-1.15^{+0.08}_{-0.06}$ & $8.8$ \\ \hline \hline
    \enddata
    \tablenotetext{\dagger}{The SMF completeness limit indicates the stellar mass limit where the incompleteness correction exceeds 10\%. }
\end{deluxetable*}

\section{Comparison Between the Cluster and Field SMFs} \label{sec:comparison_cluster_field}

We compare the MACH SMF with the field SMF as a window on the evolutionary effects of the cluster environment. The field SMF is based on SDSS field galaxies within the same redshift range as the MACH clusters. We also measure the stellar mass of the field galaxies with the same procedure we use for the galaxies in the MACH clusters (see Section \ref{sec:SDSS_MGS}). We describe the construction of field SMF in Section \ref{sec:construct_field_smf}. We compare the MACH and SDSS field SMFs in Section \ref{sec:comp_field_smf}. 

\subsection{Constructing SDSS Field SMF} \label{sec:construct_field_smf}

\begin{figure*}[hbtp]
    \centering
    \includegraphics[width=0.8\textwidth]{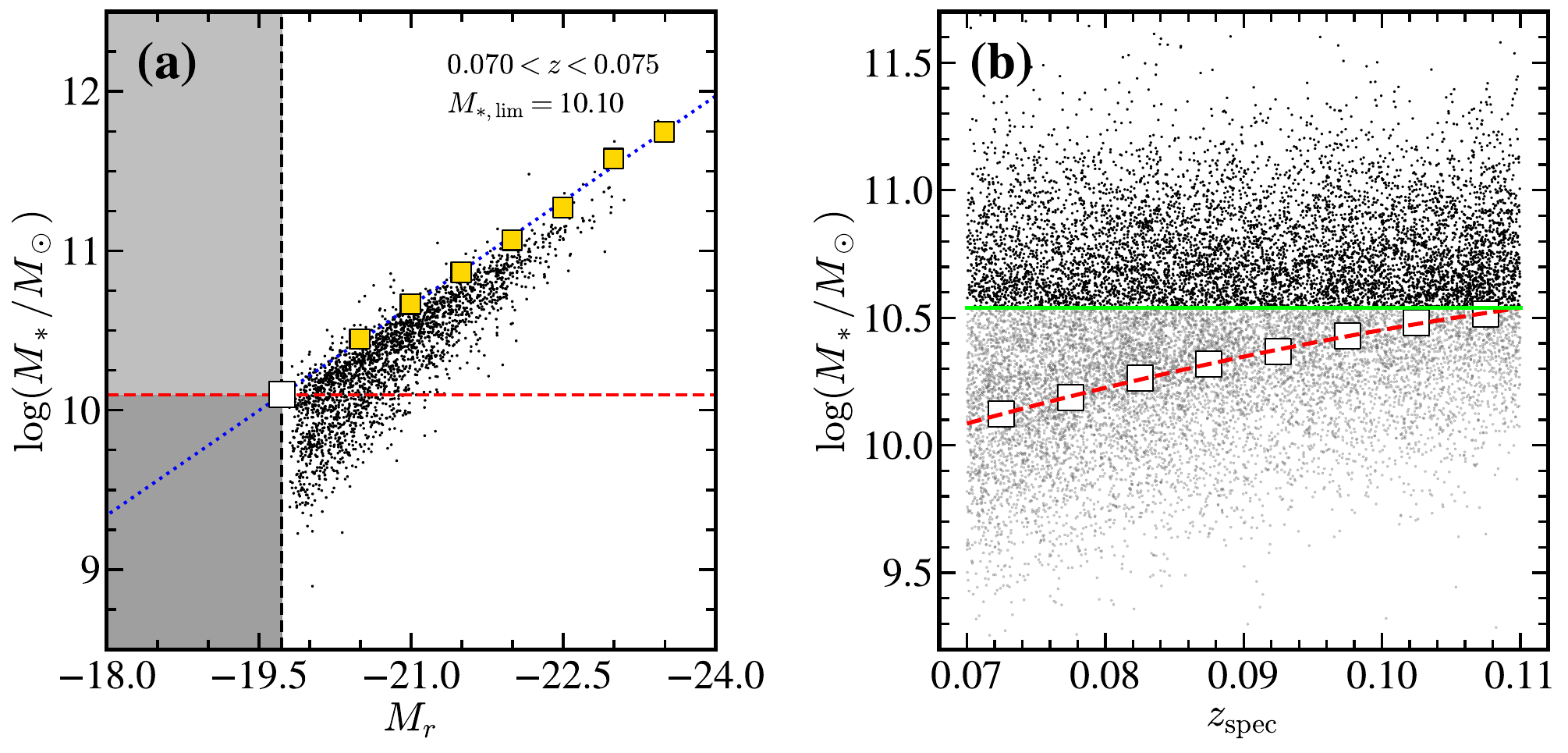}    
    \caption{(a) Stellar mass vs. $r$-band absolute magnitude for SDSS galaxies (black dots) within $0.070 < z < 0.075$. The vertical line indicates the spectroscopic survey completeness limit for the SDSS field sample ($r = 17.77$) at the redshift of this subsample. Yellow squares mark the stellar mass boundary of the top 10\% of galaxies within the absolute magnitude bins of the redshift subsample. The white square shows the stellar mass completeness limit for the redshift subsample. (b) The stellar mass completeness limit of the SDSS field sample as a function of redshift. The red dashed line indicates the best-fit second-order polynomial to the stellar mass completeness limit of each redshift subsample. Green solid line shows the stellar mass completeness limit for SDSS field sample. For clarity, we use only 10\% of the SDSS field sample.}
    \label{fig:fieldSMF_vmaxcal}
\end{figure*}

We measure the stellar mass function of SDSS field galaxies based on SDSS DR18 spectroscopy. The SDSS is a magnitude-limited survey $\sim 92.3\%$ complete to $r=17.77$ (see \citealp{Lazo2018}). However, deriving the SMF based on the magnitude-limited survey results in incompleteness of the SMF measurement due to the scatter in the stellar mass to luminosity relation. For example, if a galaxy with large stellar mass at given magnitude is excluded from the magnitude-limited sample because of the magnitude cut, the resulting stellar mass function becomes incomplete (e.g., \citealp{Weigel2016}, as we discuss further in this section). Similarly, \citet{Sohn2017b} also discuss the biases introduced in the velocity dispersion function induced by a magnitude-limited sample. We thus first construct a stellar-mass complete and volume-limited SDSS field galaxy sample.  

Figure \ref{fig:fieldSMF_vmaxcal} (a) shows the relation between stellar mass ($\rm M_{*}$) and $r$-band absolute magnitude ($M_{r}$). At a given $M_{r}$, the stellar masses of SDSS field galaxies exhibit substantial scatter ($\sim 0.5$--$1.0$ dex). In other words, converting the magnitude limit directly into a stellar mass limit results in an incomplete sample in terms of stellar mass.

We determine the stellar mass limit of the SDSS field galaxy sample following the method described in \citet{Sohn2017b}, who derive the velocity dispersion completeness limit for the same sample. Here, we apply the same technique, but for stellar mass. The detailed procedure is as follows: 

\begin{enumerate}
\item We first select redshift subsamples with $\Delta z = 0.005$ ranging from $z = 0.07 - 0.11$ (e.g., Figure \ref{fig:fieldSMF_vmaxcal} (a)). 

\item We compute the $M_{r}$ completeness limit (the black vertical line in Figure \ref{fig:fieldSMF_vmaxcal} (a)) at the mean redshift of the subsample corresponding to $r= 17.77$.

\item We determine the stellar mass boundary ($\mathrm {M_{*, 90\%}} (M_{r})$ (yellow squares in Figure \ref{fig:fieldSMF_vmaxcal} (a)) where the top 10\% of galaxies in the redshift subsample appears as a function of $M_{r}$. The blue dotted line in Figure \ref{fig:fieldSMF_vmaxcal} (a) shows the best-fit linear relation.

\item We derive the intersection between the $M_{r}$ limit (i.e., $M_{r, \rm{lim}}$) and $\mathrm{M_{*, 90\%}} (M_{r})$. The stellar mass of the intersection point corresponds to the $\rm M_{*}$ completeness limit (i.e., $\rm M_{*, \rm{lim}}$) of the redshift subsample (the red square (and the red horizontal line) in Figure \ref{fig:fieldSMF_vmaxcal} (a)). Given the spread of the stellar mass at a given absolute magnitude, $\sim10\%$ of faint galaxies below the magnitude limit (the vertical line) may have stellar masses above the stellar mass limit (the horizontal line). In other words, the SMF based on the magnitude limited sample is incomplete below the mass completeness limit. 

\item We derive the $\rm M_{*}$ limit as a function of redshift by repeating step 1 -- 4 for all redshift subsamples. The red line in Figure \ref{fig:fieldSMF_vmaxcal} (b) shows the best-fit polynomial relations for $\mathrm{ M_{*, \rm{lim}}} (z)$: 
\begin{equation}
    \label{eq:masslim}
    \log (\mathrm{ M_{*, \rm{lim}}} (z)/\Mdot) = -88.836z^2 + 27.326z + 8.608.
\end{equation}

\item We finally define the stellar-mass complete and volume-limited sample by selecting galaxies above the green horizontal line in Figure \ref{fig:fieldSMF_vmaxcal} (b) (i.e., $\logMdot > \log (\mathrm{M_{*, \rm{lim}}} (z=0.11) /\rm M_\odot) = 10.538$). 
\end{enumerate}

The final stellar mass complete-volume limited sample contains 69594 galaxies within $0.07 < z < 0.11$. The SDSS field sample still contains a large number of galaxies below the stellar mass limit, but incompleteness in the spectroscopic survey requires that we limit the field SMF to the stellar mass complete-volume limited sample.

\begin{figure}[t]
\centering
\includegraphics[width=0.45\textwidth]{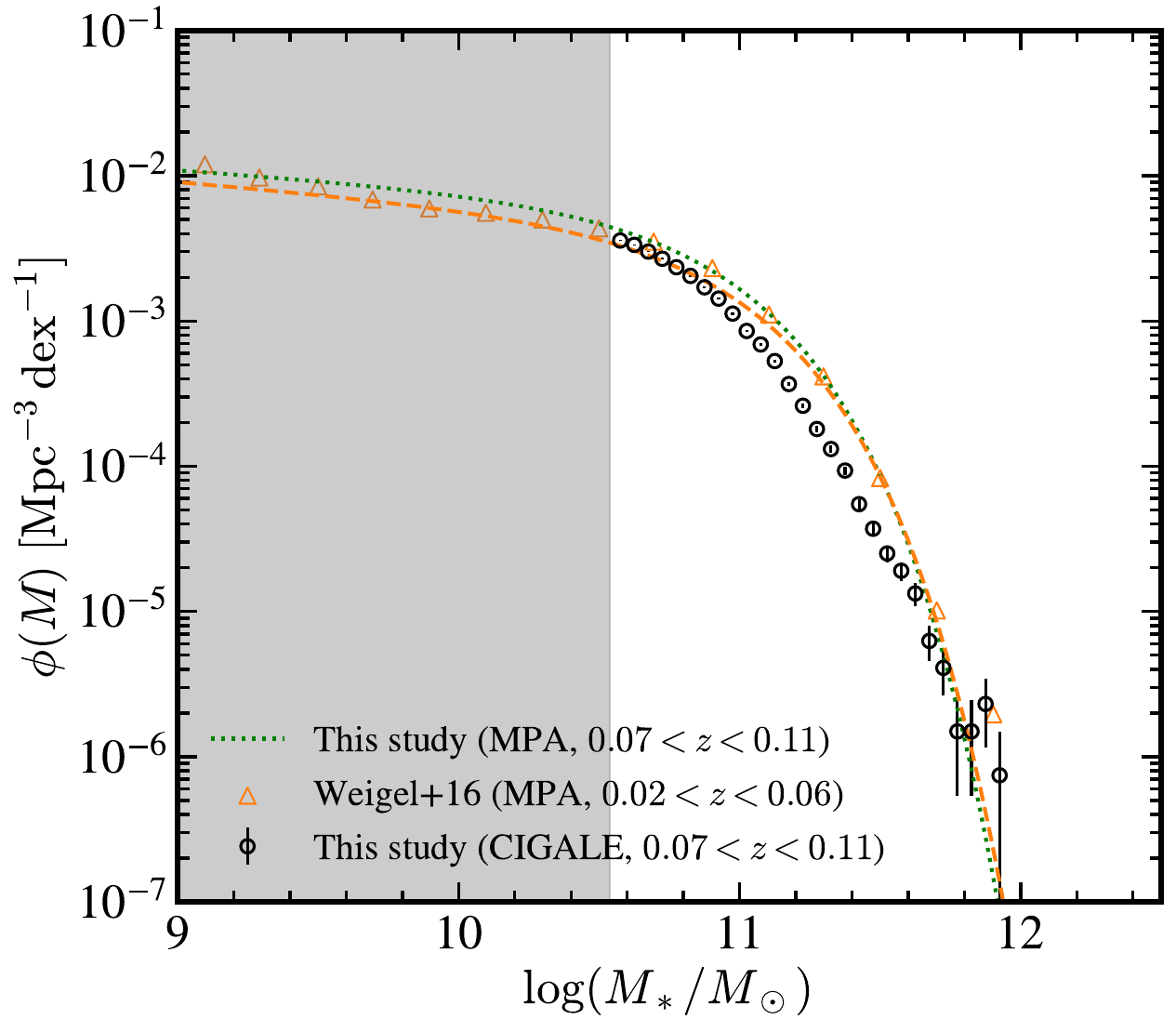}
\caption{The SMF of the SDSS field galaxies for $0.07 < z < 0.11$ (black symbols) based on CIGALE stellar mass estimates. The green dotted line shows the same SMF, but based on MPA stellar mass estimates translated from the correlation between the MPA and CIGALE stellar mass (Figure \ref{fig:Mstellar_tool_offset} (b)). For comparison, we plot the SMF based on the MPA stellar masses for SDSS field galaxies within $0.02 < z < 0.06$ \citep{Weigel2016}, shown as orange triangles with a dashed line.}
\label{fig:field_smf}
\end{figure}

Figure \ref{fig:field_smf} shows the SDSS field SMF compared with the field SMF derived from \citet{Weigel2016}. The SMF from \citet{Weigel2016} is derived from the SDSS spectroscopic sample, but based on the MPA stellar mass and the galaxies in a lower redshift range (i.e., $0.02 < z < 0.06$). The SDSS field SMF we derive differs from the SMF from \citet{Weigel2016}, particularly for $11 < \logMdot < 11.5$. 

The offset between the SMF we derive and the SMF from \citet{Weigel2016} results from the difference in the stellar mass estimates (Figure \ref{fig:field_smf}). To explore this issue, we convert the CIGALE to MPA stellar mass measurement in the SDSS field sample using the stellar mass offset between two stellar mass catalogs (see Figure \ref{fig:Mstellar_tool_offset} (b)). The green dotted line in Figure \ref{fig:field_smf} shows the field SMF for the SDSS field sample with the estimated MPA stellar mass. The difference between the SDSS field SMF with MPA stellar mass and the field SMF determined by \citet{Weigel2016} is negligible. This agreement implies that a major difference between estimates of the field SMFs probably arises from differences in stellar mass computations.

\subsection{Comparison between Field and Cluster SMFs} \label{sec:comp_field_smf}

We next compare the field and cluster SMFs to investigate the impact of the high-density environment on the stellar mass distributions. A challenge in comparing the cluster and field SMFs is the different definitions of these measurements; the cluster SMF indicates the number of cluster members within $R_{200}$ (in our case) per mass bin; the field SMF is the number of galaxies per mass bin per unit volume. Thus, a relative normalization is required for a fair comparison between the two SMFs. Here, we normalize both field and cluster SMFs by the unit mass of $10^{15}~\Mdot$, following \citet{vanderBurg2018}. 

We first compute the total mass included in the comoving volume of the SDSS field sample to normalize the field SMF. The total mass within the volume covered by the SDSS field sample is:
\begin{equation}
M_{\rm total} = \Omega_m \rho_{\text{crit},0} \int_{z_1}^{z_2}  dV_c  \sim 2.368 \times 10^{18}~\Mdot, 
\end{equation}
where $z_{1} = 0.07$ and $z_{2} = 0.11$ for the SDSS field sample, and $dV_c$ is the comoving volume (see Appendix \ref{sec:appendix_field_mass_derivation} for details). We then normalize the field SMF to have the galaxy number count per mass bin per $10^{15}~\Mdot$. 

We next normalize the cluster SMF in the same units as the normalized field SMF. The cluster SMFs we measure include member galaxies within a projected radius (i.e., either $R_{500}$ or $R_{200}$). For simplicity, we assume that all cluster members within a projected radius are located within a spherical volume of that radius. We then assume that the $M_{200}$ and $M_{500}$ derived from the caustic technique indicates the total mass within the spherical volume limited by $R_{200}$ or $R_{500}$. We stack all of the MACH SMFs and divide by the sum of $M_{500}$ or $M_{200}$ for all of the MACH clusters. 

\begin{figure}[t]
\centering
\includegraphics[width=0.45\textwidth]{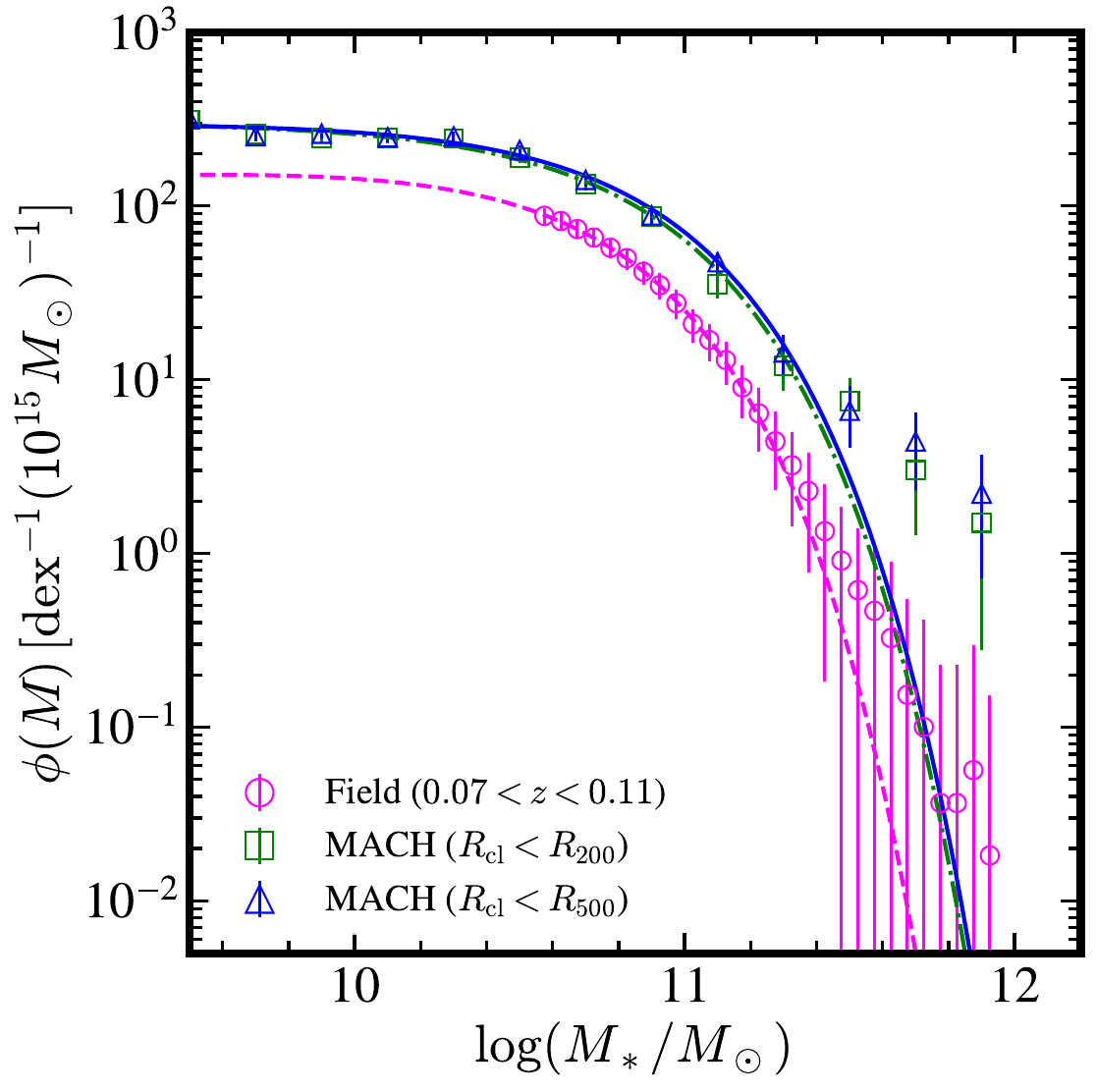}
\caption{Field and cluster SMFs normalized at $10^{15}~\Mdot$. We use CIGALE stellar masses for both field and cluster samples. Magenta circles show the SDSS field SMF. Green squares and blue triangles display the stacked MACH SMF within $R_{200}$ and $R_{500}$, respectively. Lines show the best-fit Schechter function for each SMF.}
\label{fig:FieldClusterComp}    
\end{figure}

Figure \ref{fig:FieldClusterComp} compares the normalized field (magenta circles) and cluster SMFs measured within $R_{500}$ (blue triangles) and $R_{200}$ (green squares). The cluster SMFs measured within $R_{500}$ and $R_{200}$ are similar to each other. 

The cluster SMF clearly exceeds the field SMF across the entire mass range we explore. The excess in the cluster SMF relative to the field SMF is also clear in the work by $z>1$ \citep{vanderBurg2013}. The excess varies with stellar mass. In the $10.5 < \logMdot < 11.4$ range, the cluster SMF is about twice as large as the field SMF. For $\logMdot > 11.4$, the excess is even greater; massive galaxies are more abundant in cluster environments.

\section{DISCUSSION} \label{sec:discussion}

We derive SMFs for nine massive clusters based on extremely dense spectroscopic surveys. These MACH SMFs follow a single Schechter function with an excess at $\logMdot \gtrsim 11$. Comparison between the MACH and field SMFs suggests that massive galaxies with $\logMdot \gtrsim 11$ are located preferentially in cluster environments. Here, we explore the dependence of  cluster SMFs on the galaxy populations (i.e., quiescent vs. star-forming) in Section \ref{sec:populations}. We then compare cluster SMFs with SMFs derived from cosmological simulation in Section \ref{sec:sim_comparison}. 

\subsection{SMFs for Quiescent and star-forming Galaxies in MACH Clusters} \label{sec:populations}

We next explore the way cluster SMFs depend on the galaxy populations (i.e., quiescent vs. star-forming). A key advantage of the MACH survey is that it is a magnitude-limited survey conducted without any color-based selection. The design of the MACH survey allows us to derive SMFs for both quiescent and star-forming populations to comparable depths, without concern for potential selection biases.

We examine the MACH SMFs separately for the quiescent and star-forming population. We classify the quiescent and star-forming galaxies based on $\dn$ (see Section \ref{sec:dn4000}). Comparing the SMFs of these two populations allows us to investigate the quenching efficiency as a function of galaxy mass in the cluster environment.

\begin{figure*}[htbp]
\centering
\includegraphics[width=\textwidth]{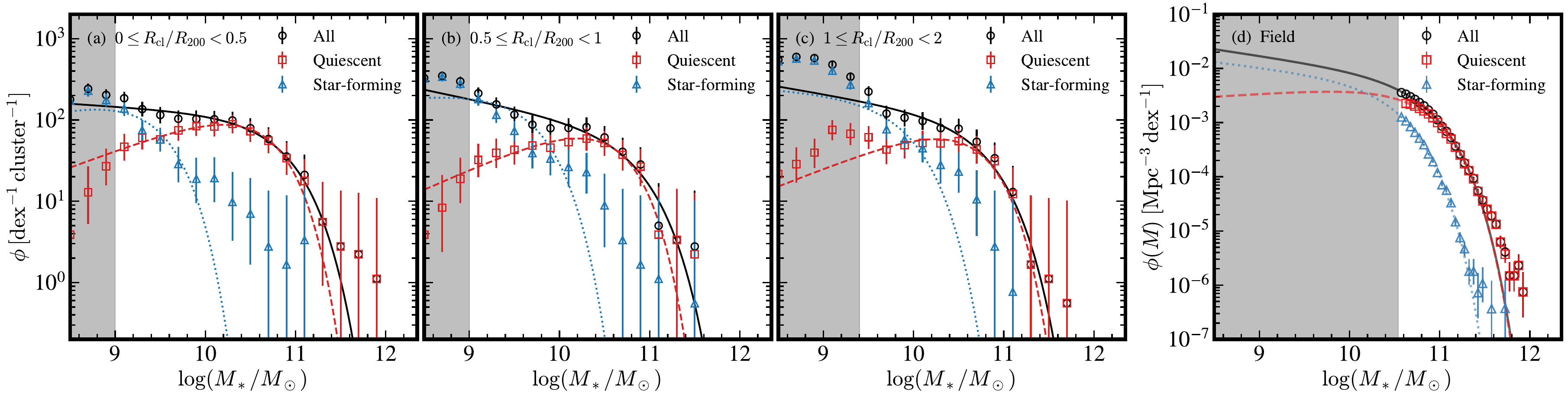}
\caption{Cluster SMFs for quiescent and star-forming populations in three radial bins: (a) $0 \leq R_\mathrm{cl}/R_{200} < 0.5$, (b) $0.5 \leq R_\mathrm{cl}/R_{200} < 1$, and (c) $1 \leq R_\mathrm{cl}/R_{200} < 2$. Black circles, red squares, and blue triangles show SMF for the entire galaxy sample, quiescent ($\dn > 1.5$) and star-forming ($\dn < 1.5$) galaxies, respectively. Lines with the same colors show the best Schechter function fit for each SMF. The gray shaded regions show the mass range where the spectroscopy is incomplete. (d) Same as (a) - (c), but for SDSS field galaxy sample.}
\label{fig:SMF_sf_q_compare}
\end{figure*}

\begin{figure}[htbp]
    \centering
    \includegraphics[width=0.45\textwidth]{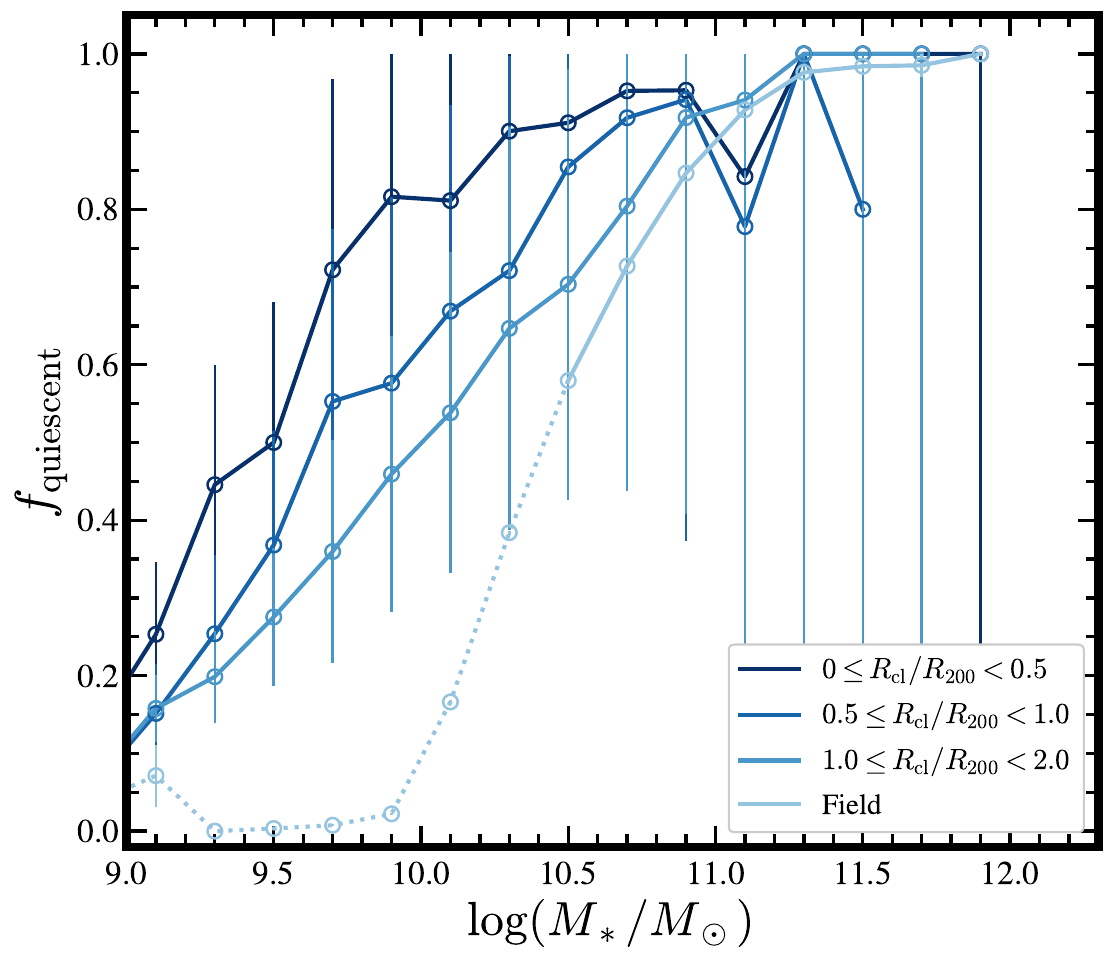}
    \caption{ The quiescent galaxy fraction ($f_{\mathrm{quiescent}}$) as a function of stellar mass in three radial bins, corresponding to those shown in Figure \ref{fig:SMF_sf_q_compare}. Darker symbols represent measurements at smaller cluster-centric radii. Light blue symbols show $f_{\mathrm{quiescent}}$ for field galaxies, and the dotted line marks the stellar mass range where the SDSS field sample becomes incomplete. Error bars indicate Poisson uncertainties. }
    \label{fig:SMF_fq_compare}
\end{figure}

Figure \ref{fig:SMF_sf_q_compare} shows the SMFs of the MACH cluster members, segregated by $\dn$; black circles mark the total SMF, red squares mark the quiescent population, and blue triangles mark star-forming population. Solid lines in the same colors indicate the best-fit Schechter functions for each SMF. Panels (a) to (d) display the mean MACH SMFs within three radial bins along with the field: (a) $R_\mathrm{cl} < 0.5R_{200}$, (b) $0.5R_{200} < R_\mathrm{cl} < R_{200}$, (c) $R_{200} < R_\mathrm{cl} < 2R_{200}$, and (d) field. 

The SMF for quiescent galaxies generally peaks at $\logMdot \sim 10.5$; the exact location of the peak varies slightly depending on the radial region we explore. The quiescent SMFs show a clear decline for $\logMdot \lesssim 10.5$. In contrast with the quiescent population, the star-forming SMF increases monotonically as the stellar mass decreases. 

We also note that the amplitude of SMFs varies differently depending on the galaxy populations. The amplitude of quiescent SMFs is higher near the cluster core. In contrast, the amplitude of star-forming SMFs is lower in the cluster core. The radial dependence of the SMF amplitudes show that the fraction of quiescent galaxies increases toward the cluster center. Star formation has ceased in a larger fraction of low mass galaxies near the cluster core. 

The fraction of quiescent galaxies ($f_{\mathrm{quiescent}}$) increases toward the cluster center at given mass. Figure \ref{fig:SMF_fq_compare} shows the $f_{\mathrm{quiescent}}$ in MACH clusters as a function of stellar mass at different projected cluster-centric distances (i.e., at $(R_\mathrm{cl}/R_{200}) < 0.5$, $0.5 < (R_\mathrm{cl}/R_{200}) < 1.0$, and $1.0 < (R_\mathrm{cl}/R_{200}) < 2.0$). For comparison, we plot the $f_{\mathrm{quiescent}}$ for the SDSS field sample. For $\logMdot>10.5$, the $f_{\mathrm{quiescent}}$ has little dependence on the environment. A noticeable change occurs at $9.5 < \logMdot < 10.5$; the $f_{\mathrm{quiescent}}$ in the cluster core is much higher than in the field and in the cluster outskirts. This change of $f_{\mathrm{quiescent}}$ in the MACH sample reflects the known mass and environmental dependent quenching processes (e.g., \citealp{Peng2010}). 

We compare the quiescent galaxy fractions in MACH clusters with those reported for other clusters and groups in the literature. For example, \citet{Deshev2017} and \citet{Sohn2019b} examined the mean $\dn$ as a function of projected cluster-centric distance for galaxies in A520 and A2029, respectively. In both clusters, the mean $\dn$ decreases systematically with increasing projected distance from the cluster center, indicating a declining quiescent galaxy fraction toward the cluster outskirts. Similarly, studies of quiescent galaxy fractions based on $\rm{H\alpha}$ emission line strengths using much larger statistical samples from SDSS or GAMA (e.g., \citealp{Wetzel2012, Davies2019, Baxter2021}) also find that the quiescent galaxy fraction is significantly smaller in the cluster outskirts than in the cluster cores. \citet{Morgan2025} reported a similar radial dependence for the Virgo cluster galaxies based on the H$_{\alpha}$ selection, finding that the quiescent fraction decreases to $\sim 50\%$ at $R_{200}$. At this radius, the quiescent fraction of Virgo cluster galaxies with ${\logMdot \sim 9}$ is comparable to that measured for MACH clusters.

\subsection{Comparison with the SMF of Simulated Clusters} \label{sec:sim_comparison}

We  next compare the observed SMFs with SMFs derived from clusters in IllustrisTNG-300 simulation \citep{Pillepich2018a, Springel2018, Nelson2018, Naiman2018, Marinacci2018}. IllustrisTNG-300 is a set of cosmological magnetohydrodynamic simulations covering a (300 Mpc)$^{3}$ box. Among the TNG simulations covering the same box size, we use IllustrisTNG300-1 (hereafter, TNG300) with the highest mass resolution (i.e., the stellar mass resolution $\sim 1.1 \times 10^{7}~\Mdot$). We use the snapshot 93, corresponding to $z = 0.07$, comparable with the redshift of MACH clusters. TNG300 provides a cluster halo catalog built based on a Friends-of-Friends algorithm. From the cluster halo catalog, we select nine clusters with $M_{200} > 5 \times 10^{14}~\Mdot$, covering the same mass range as the MACH clusters. 

We derive the SMFs of the member galaxies in the simulated clusters. For a direct comparison with observations, we select member galaxies within $R_{200}$ of each TNG300 cluster. We also obtain the stellar masses (i.e., \texttt{SubhaloStellarPhotometricsMassInRad}, the sum of member stellar particles, for cluster members. 

\begin{figure}[t]
    \centering
    \includegraphics[width=0.45\textwidth]{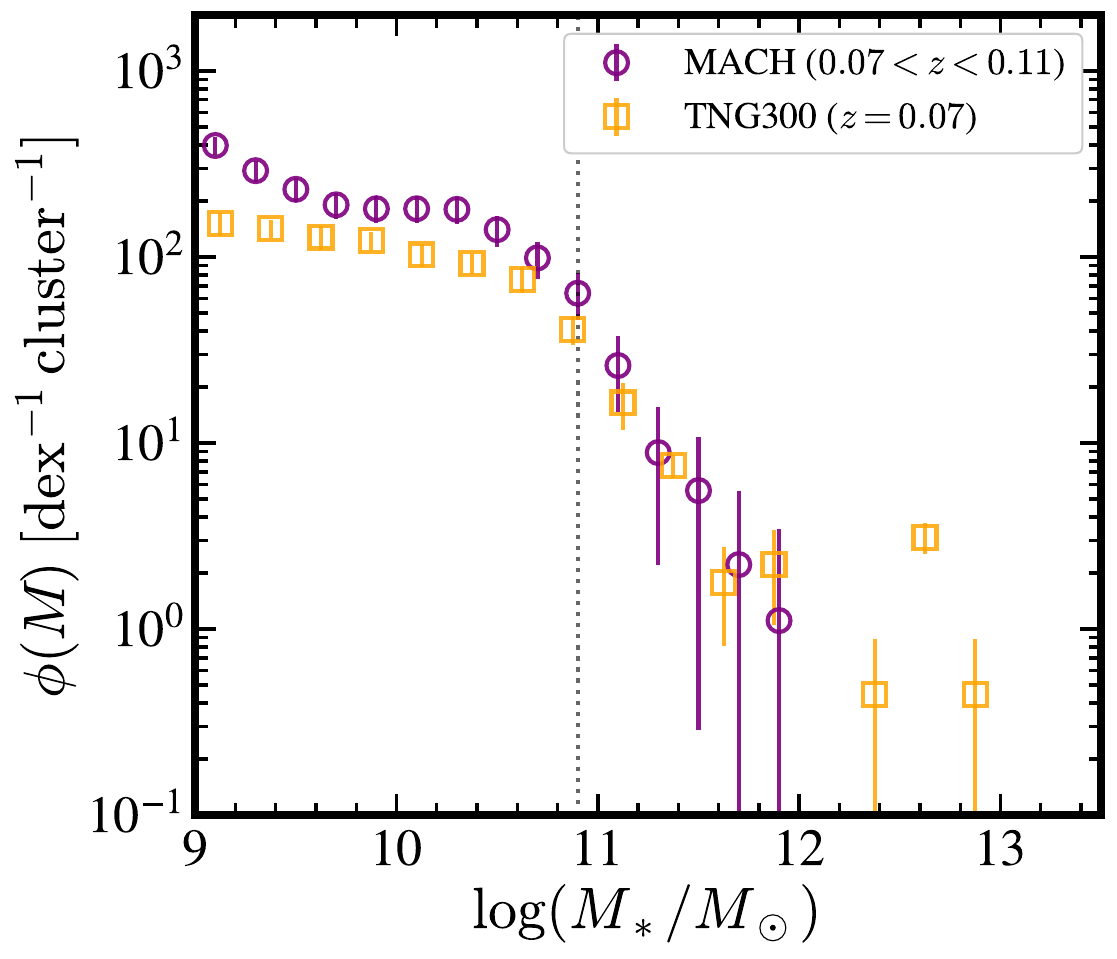}
        \caption{Comparison between the mean MACH SMF (i.e., observed, purple circles) and mean TNG cluster SMF (i.e., simulated, orange squares).} 
        \label{fig:obs_sim_compare}
\end{figure}

Figure~\ref{fig:obs_sim_compare} compares the observed and simulated SMFs. We display the mean SMFs for nine MACH SMFs and nine TNG SMFs. Purple circles and orange squares display the observed MACH SMFs and the TNG cluster SMFs, respectively. The observed and simulated SMF match well over the range $10.5 < \logMdot < 11.6$.

At the high-mass end ($\logMdot > 11.6$), the observed and simulated SMFs differ slightly. Although the number of massive galaxies is small, the brightest cluster galaxies (BCGs) in the simulations are significantly more massive than those in the observed clusters. This discrepancy may arise from difficulties in separating stellar particles belonging to the BCGs from those associated with the extended cluster halos.

Another noticeable difference appears at the low-mass end ($\logMdot < 10.5$), where the observed SMF exceeds the simulated SMF significantly. The observed cluster sample contains $\sim 90\%$ more galaxies than the simulated cluster samples within $9 < \logMdot < 10.5$. The significant offset suggests that galaxy formation in the TNG simulations is less efficient at low mass compared with the observations. \citet{Sohn2024} also reported that TNG300 clusters at $z = 0$ contain fewer member galaxies than observed clusters (see their Figure 2).

The MACH SMFs demonstrate that complete spectroscopic samples provide valuable constraints for calibrating galaxy formation models in cosmological simulations. Upcoming large-scale photometric and spectroscopic surveys, including LSST, DESI, Subaru/PFS, MOONS, and 4MOST, will deliver extensive spectroscopic datasets for both cluster and field galaxies. Comprehensive comparisons between observed and simulated SMFs will enable a deeper understanding of the physical processes governing galaxy formation. 

\section{CONCLUSION} \label{sec:conclusion}

We measure stellar mass functions (SMFs) for nine most massive galaxy clusters in the local universe ($0.07 < z < 0.11$) based on the MAssive Cluster survey with Hectospec (MACH). The MACH survey is a deep, dense, complete, and homogeneous spectroscopic survey. We construct robust samples of cluster members using the caustic technique and derive the physical properties of their members, including stellar mass and $\dn$.

We derive MACH SMFs to a limiting $\logMdot \sim 8.5$. The MACH SMFs are complete for $\logMdot > 9$ without applying any correction for spectroscopic incompleteness. To extend the SMF measurements to lower masses $\logMdot \sim 8.5$, we apply an empirical correction that accounts for spectroscopic incompleteness as a function of magnitude, color, and cluster-centric distance.

We compare the MACH SMFs with a carefully constructed field SMF derived from galaxies within the same redshift range ($0.07 < z < 0.11$) using SDSS DR18 spectroscopy. Because SDSS spectroscopy is shallower than MACH, the SDSS field SMF is complete only for $\logMdot \gtrsim 10.5$. The  shape of the field SMF is consistent with the MACH SMF for $10.5 < \logMdot < 11.4$, but with an amplitude smaller by a factor of $\sim 2$. For $\logMdot > 11.4$, the MACH SMFs show a clear excess relative to the field SMF, probably driven by the significant contribution from massive galaxies in dense cluster environments.

The complete MACH spectroscopy without color selection enables homogeneous measurements of SMFs for both quiescent and star-forming populations to comparable depths. The two populations exhibit distinct SMF shapes; the quiescent SMFs display a curved profile with a peak at $\logMdot \sim 10.5$ and the star-forming SMFs increase monotonically toward lower stellar masses. We also examine the radial dependence of these SMFs. The peak of the quiescent SMF shifts toward lower stellar mass in the cluster core, indicating that a larger fraction of low-mass galaxies are quenched in the central regions. Similarly, the quiescent fraction increases with decreasing stellar mass toward the center.

We further compare the mean MACH SMF with the SMF of comparably massive clusters from the Illustris-TNG300 simulation. The two agree well for $10.5 < \logMdot < 11.4$. At higher stellar masses ($\logMdot > 11.4$), simulated clusters host more massive BCGs than the observed clusters. This difference may arise from issues in separating stellar particles belonging to the BCGs from those associated with the surrounding cluster halos. For $\logMdot < 10.5$, the MACH clusters contain roughly a factor of two more galaxies than the  analogous simulated clusters. This comparison demonstrates that observed cluster SMFs derived from complete spectroscopy provide a basis for testing and refining galaxy formation models in cosmological simulations. 

The SMF is a widely used tool for probing galaxy formation efficiency as a function of galaxy mass, environment, and the properties of underlying dark matter halos. The SMFs derived from the deep and dense spectroscopy of nine massive clusters provide a foundation for comparing results from the field, high-redshift clusters, and cosmological simulations. Future dense spectroscopic surveys of massive clusters (e.g., Subaru/PFS, 4MOST, WST) will yield even deeper datasets, enabling a more detailed understanding of the stellar mass evolution of galaxies in dense environments. 

\section*{Acknowledgements}
We first thank Perry Berlind and Michael Calkins for operating Hectospec. We thank Susan Tokarz, Jaehyon Rhee, Sean Moran, Warren Brown, and Nelson Caldwell for their tremendous contributions to the observing preparation and data reduction.  This work was supported by the National Research Foundation of Korea (NRF) grant funded by the Korea government (MSIT) (RS-2023-00210597). This work was also supported by the Global-LAMP Program of the National Research Foundation of Korea (NRF) grant funded by the Ministry of Education (No. RS-2023-00301976). A.D. acknowledges partial support from the INFN grant InDark. 

Funding for the Sloan Digital Sky Survey IV has been provided by the Alfred P. Sloan Foundation, the U.S. Department of Energy Office of Science, and the Participating Institutions. SDSS-IV acknowledges support and resources from the Center for High Performance Computing at the University of Utah. The SDSS website is www.sdss.org. SDSS-IV is managed by the Astrophysical Research Consortium for the Participating Institutions of the SDSS Collaboration including the Brazilian Participation Group, the Carnegie Institution for Science, Carnegie Mellon University, Center for Astrophysics—Harvard \& Smithsonian, the Chilean Participation Group, the French Participation Group, Instituto de Astrof\'isica de Canarias, The Johns Hopkins University, Kavli Institute for the Physics and Mathematics of the Universe (IPMU)/University of Tokyo, the Korean Participation Group, Lawrence Berkeley National Laboratory, Leibniz Institut f\"ur Astrophysik Potsdam (AIP), Max-Planck-Institut für Astronomie (MPIA Heidelberg), Max-Planck-Institut f\"ur Astrophysik (MPA Garching), Max-Planck-Institut für Extraterrestrische Physik (MPE), National Astronomical Observatories of China, New Mexico State University, New York University, University of Notre Dame, Observatário Nacional/MCTI, The Ohio State University, Pennsylvania State University, Shanghai Astronomical Observatory, United Kingdom Participation Group, Universidad Nacional Autónoma de México, University of Arizona, University of Colorado Boulder, University of Oxford, University of Portsmouth, University of Utah, University of Virginia, University of Washington, University of Wisconsin, Vanderbilt University, and Yale University.

\newpage 

\appendix  
\section{Derivation of the total mass in the volume covered by the SDSS field sample} \label{sec:appendix_field_mass_derivation}

We compute the total matter mass included in the comoving volume within $z_{1} < z < z_{2}$ for the field sample as follows.
\begin{align*}
M_\mathrm{matter}(z_1, z_2) &= \int_{z_1}^{z_2} \int \rho_m(z) \, dV_{\text{phys}} \\
&= \int_{z_1}^{z_2} \int \Omega_m \rho_{\text{crit},0} (1+z)^3 \, \frac{dV_c}{(1+z)^3} \\
&= \Omega_m \rho_{\text{crit},0} \int_{z_1}^{z_2} \int dV_c . \\ 
&= \Omega_m \rho_{\mathrm{crit},0} \, 4 \pi f_{\mathrm{survey}} 
\int_{z_1}^{z_2} \frac{c}{H_0} \frac{D_C(z)^2}{E(z)} \, dz \quad  (\because dV_c = \frac{c}{H_0} \frac{D_C(z)^2}{E(z)} \, d\Omega \, dz. ) \\
&= \Omega_m \rho_{\mathrm{crit},0} 
\frac{4 \pi f_{\mathrm{survey}}}{3} 
\left[ D_C(z_2)^3 - D_C(z_1)^3 \right] \\ 
&= \Omega_m  \cdot \frac{3H_0^2}{8\pi G} 
\frac{4 \pi f_{\mathrm{survey}}}{3} 
\left[ D_C(z_2)^3 - D_C(z_1)^3 \right] \quad (\text{where, } \, D_c(z) = \frac{c}{H_0} \int_{0}^{z} \frac{dz'}{\sqrt{\Omega_m (1+z')^3 + \Omega_\Lambda}}). 
\end{align*}
Here, $f_{\text{survey}}$ is the ratio between the solid angle covered by the SDSS field sample and the solid angle of the entire sky. 

\section{NASA/IPAC Extragalactic Database (NED) Redshifts of galaxies in the MACH cluster field} \label{sec:NED_database}

\begin{deluxetable*}{cccccccc}[h]
\tablewidth{\textwidth}
\label{tab:NED_redshift}
\setlength{\tabcolsep}{7pt}
\renewcommand{\arraystretch}{1.1}
\tabletypesize{\normalsize}
\renewcommand{\tablenotemark}[1]{\textsuperscript{\small#1}}
\caption{Spectroscopic Redshifts for MACH Cluster Galaxies within $3R_{200}$ from NED}
\tablehead{
\colhead{ID} & \colhead{SDSS Objid} & \colhead{R.A. (deg)} & \colhead{Dec. (deg)} & \colhead{Cluster ID\tablenotemark{a}} & \colhead{$z$} & \colhead{$e_z$\tablenotemark{b}} & \colhead{$z$ Reference \tablenotemark{c}}
}
\startdata
1 & 1237662697033367784 & 229.575651 & 28.397926 & A2065 & 0.073794 & 0.000203 & 15 \\
2 & 1237662697033433433 & 229.749189 & 28.354935 & A2065 & 0.117420 & 0.000060 & 4 \\
3 & 1237662697033433453 & 229.761652 & 28.342230 & A2065 & 0.074760 & 0.000170 & 4 \\
4 & 1237665103827763362 & 229.919206 & 27.585193 & A2065 & 0.122700 & 0.000100 & 10 \\
5 & 1237664853113176095 & 229.940686 & 28.492095 & A2065 & 0.075836 & 0.000130 & 15 \\
6 & 1237664853649981673 & 230.025424 & 28.796036 & A2065 & 0.073650 & 0.000090 & 4 \\
7 & 1237662697033695440 & 230.276890 & 28.112214 & A2065 & 0.084125 & 0.000500 & 25 \\ \hline \hline
\enddata
\tablenotetext{a}{Cluster ID where galaxies are located in.}
\tablenotetext{b}{The redshift error is set to $0.0001$ when the redshift error is not available from NED.}
\tablenotetext{c}{
(1) \citet{Sohn2020}, (2) \citet{Golovich2019}, (3) \citet{Sohn2019a},
(4) \citet{Keeney2018}, (5) \citet{Lin2018}, (6) \citet{Sohn2017},
(7) \citet{Lee2017}, (8) \citet{Rines2016}, (9) \citet{Albareti2016},
(10) \citet{Wen2015}, (11) \citet{Drinkwater2010}, (12) \citet{Cava2009},
(13) \citet{Springob2005}, (14) \citet{Abazajian2004}, (15) \citet{Miller2002},
(16) \citet{Wegner2001}, (17) \citet{Falco1999},
(18) \citet{Huchra1999}, (19) \citet{Maurogordato1997},
(20) \citet{Small1997}, (21) \citet{Szomoru1994}, (22) \citet{Fouque1992catalogue},
(23) \citet{Batuski1991}, (24) \citet{Karachentsev1990}, (25) \citet{Postman1988},
(26) \citet{Humason1956}.
}
\tablecomments{A full version of this table is provided in machine‐readable format in the online journal. The portion shown here serves as an example of its structure and content.}
\end{deluxetable*}


\bibliographystyle{aasjournal}
\bibliography{ms}{}



\end{document}